\documentclass[12pt]{report}
\usepackage{hyperref}
\usepackage{amsmath}
\usepackage{mathrsfs}
\usepackage{cite}
\usepackage{subfloat}
\usepackage{xcolor}
\usepackage{adjustbox}
\usepackage{caption}
\usepackage{subcaption}
\usepackage{graphicx}
\usepackage{relsize}
\usepackage{amssymb}
\usepackage{float}
\usepackage[utf8]{inputenc}
\usepackage{appendix}
\usepackage{xcolor}
\usepackage[bottom=1in,top=1.2in]{geometry}
\usepackage{setspace} 
\usepackage{imakeidx}
\usepackage{subcaption}
\usepackage{mathtools}

\DeclareUnicodeCharacter{2212}{---}
\usepackage{authblk}
\begin{document}
\title{\textbf{E. C. G. Sudarshan and his Diagonal Representation in Quantum Optics}}
\author[1]{Sobhan K. Sounda \thanks{sobhan.physics@presiuniv.ac.in}}
\author[2]{Koushik Mandal \thanks{koushik.mandal@iopb.res.in}}
\affil[1]{Department of Physics, Presidency University, 86/1 College Street, Kolkata-700073, India}
\affil[2]{S. N. Bose National Centre for Basic Sciences, Kolkata-700106, India}
\affil[2]{Institute of Physics, Sachivalaya Marg, Bhubaneswar-751005, India}
\date{}
\maketitle
\begin{abstract}
The essential of this monograph is to reproduce the results of Sudarshan's paper ``Equivalence of semiclassical and quantum mechanical descriptions of statistical light beams''[\textbf{Phys. Rev. Lett. 10, 277(1963).}] published in 1963. To go in that direction we also describe the linear harmonic oscillator and its algebraic equivalence with a single-mode quantized beam of light. Coherent states and their over completeness properties along with the diagonal representation of the density matrix are essential as a prerequisite to reproduce the result. Most of the textbooks meant for graduate students describe linear harmonic oscillators in great detail but they cryptically mention coherent states and their associated properties. So it is difficult for the students to understand the essence of Sudarshan's paper which contains only seven equations. This monograph is written in such a lucid manner that without the help of experts one can understand it.
\end{abstract}

\pagenumbering{roman}
\newpage
\tableofcontents
\newpage
\chapter*{Preface} 
The two and a half pages paper Equivalence of semiclassical and quantum mechanical descriptions of statistical light beams published in Physical Review Letters
in 1963 bears the signature of E. C. G. Sudarshan’s adventure in Non-classical optics. It is considered as a path
breaking paper in Theoretical Quantum Optics which provides a quantitative method to demarcate classical and non
classical light.\\
Most of the graduate students in Physics and
theoretical physicist unaware of this path breaking but it is to be easily understood from the knowledge of basic quantum mechanics. The essential of this monograph is to reproduce the results of Sudarshan’s paper. To go in that direction we also describe the linear harmonic oscillator and its algebraic equivalence with a single mode quantized beam of light, coherent states and it’s over completeness properties, diagonal representation of density matrix as a prerequisite. Most of the text books meant for graduate students describe linear harmonic oscillator in great detail but they cryptically mention coherent states and its associated properties. So it is difficult for the students to understand the essence of Sudarshan’s paper which contains only seven equations. This monograph is written in such a lucid manner that without help of experts one can understand it. Why one should read E.C.G’s paper? In this regard  we quote Prof. Jeff Kimble \textit{...By now, a zoology of non classical sources shine in laboratories around the world, with applications ranging from quantum measurement to quantum computation and communication. The Optical Equivalence Theorem of Sudarshan has been the critical, guiding light in these quests.} The authors of this monograph hope that this mere ninety pages material will help to understand E. C. G’s paper in detail and may fill the gap between the Physics community and Sudarshan’s Pioneering work.\\
\begin{flushright}
Sobhan K. Sounda\\
Koushik Mandal\\
Kolkata, India
\end{flushright}
\newpage
\pagenumbering{arabic}
\chapter{Introduction}
\vspace{0.3cm}
E. C. G. Sudarshan is known to be the most gifted theoretical physicist of Indian origin in the second half of 20th century. Of all his work in different fields, two of E.C.G's achievements are highly regarded. These are the V-A theory and Diagonal Representation or Optical Equivalence theorem. These two achievements are considered that they are of the Nobel class. In an article\cite{Simon} Prof. Raja Simon of IMSC, Chennai remarked that \emph {...if the V-A theory represents Professor Sudarshan's mid-century adventure in Particle Physics, the Diagonal Representation or Optical Equivalence Theorem represents his mid-century adventure in Non-classical Optics.}\\
E. C. G's epoch making contribution in the field of Quantum Optics was beautifully summarized by Prof. Jeff Kimble in a conference which was named after \emph{Sudarshan: Seven Science Quests}. Now we quote from the abstract that was presented in the  conference by Prof. Kimble:\emph{ In 1963 Prof. E. C. G. Sudarshan presented the Optical Equivalence Theorem and thereby provided a quantitative, model independent boundary between the classical and manifestly quantum domains of light. More than a decade then passed before non-classical light first flickered dimly in the laboratory. By now, a zoology of non-classical sources shine in laboratories around the world, with applications ranging from quantum measurement to quantum computation and communication. The Optical Equivalence Theorem of Sudarshan has been the critical guiding light in these quests.}\\

In this monograph we take the liberty to describe the remarkable discovery of the Diagonal Coherent State Representation and the Optical Equivalence Theorem by following E. C. G's paper:\emph{Equivalence of semi-classical and quantum mechanical descriptions of statistical light beams} \cite{E.C.G} which is central to quantum optics. \\\\
A review of linear harmonic oscillator(LHO) and its algebraic equivalence with quantized electromagnetic beam is presented for the reader. Some properties of coherent states and particular emphasis to over-completeness which led to the discovery of Diagonal Coherent State Representation and the Optical Equivalence Theorem will be discussed. Finally we discuss some phenomena like photon anti-bunching which can be well understood using E. C. G.'s diagonal representation and quasi-probability distribution.
\chapter{Harmonic Oscillator}
\section{Coherent state and harmonic oscillator}
\vspace*{0.3cm}
In 1926 Schrodinger\cite{ESch} first proposed the idea of `harmonic oscillator coherent states' in the context of transition from quantum to classical physics. He referred to these classical states as the state of minimum uncertainty product. After more than three decades properties of coherent states were further investigated by Klauder and Bargmann\cite{Klauder,Bargmann}. Coherent states came to center stage of Quantum optics  with the seminal paper: \emph{Photon Correlations} by  R. J. Glauber\cite{Glauber6} in 1963 and  his  two other paper\cite{Glauber7,Glauber8} published in the same year. Glauber described coherent state as the eigen states of the annihilation operator. This was done in the context of electromagnetic coherence functions which was emerging then as a great importance in quantum optics. Here, in this chapter we introduce linear harmonic oscillator in one dimension and then we will carry it to describe coherent state.
\section{Linear harmonic oscillator in  one dimension}
 \vspace{0.3cm}
 In classical mechanics, the simplest model of harmonic oscillator is a mass $m$ attached to a spring of stiffness constant $k$. If we ignore friction, then for small displacement of the mass along $X-$ axis, its potential energy will be $V(x) = \frac{1}{2} k x^{2}$. It has parabolic form and is symmetric about the origin i.e. $x = 0$. We know that the frequency of oscillation is $\omega = \sqrt\frac{k}{m}$. Therefore, the form of the potential energy is $V(x) = \frac{1}{2}m\omega^{2} x^{2}$ and  classical Hamiltonian of the spring-mass system is 
 \begin{equation}
 H_{cl} = \frac{p_{x}^2}{2m} + \frac{1}{2}m\omega^{2}x^{2}
 \end{equation}
 Now we can turn it into quantum mechanics simply by replacing classical canonical conjugate variables by quantum mechanical hermitian operators expressed in position basis like: $p_{x} \rightarrow \hat{p}_x\equiv - i\hbar\frac{\partial}{\partial x}$ and $x \rightarrow \hat{x}\equiv x$. Thus, the quantum mechanical Hamiltonian
 \begin{equation}
 \hat{H}_{qm} = \frac{\hat{p}_{x}^2}{2m} + \frac{1}{2}m\omega^2\hat{x}^2 
 \end{equation}
 takes the form in position basis as
 \begin{equation}
 H_{qm}= - \frac{{\hbar}^2}{2m}\frac{\partial^2} {\partial x^2} + \frac{1}{2}m\omega^2x^2
 \end{equation}
 Now, the time dependent Schrodinger equation is
 \begin{equation}
 i\hbar\frac{\partial}{\partial t}\psi(x,t) = \left(-\frac{{\hbar}^2}{2m}\frac{\partial ^2} {\partial x^2} + \frac{1}{2}m\omega^2x^2\right)\psi (x,t)
 \end{equation}
 Here, we do the separation of variables as $\psi (x,t) = \phi (x) T(t)$. On
 substituting $\psi (x,t)$ in time dependent Schrodinger equation we obtain \\
 \begin{eqnarray}
 \frac{1}{\phi(x)}\left[-\frac{\hbar^2}{2m}\frac{d ^2} {d x^2}+\frac{1}{2}m\omega^2x^2\right]\phi (x) =  \frac{i\hbar}{T}\frac{d}{d t}T(t) =\lambda=E 
 \end{eqnarray}
 In the above equation, left hand side is a function of space coordinate only and that on the right hand side is a function of time coordinate only. So both sides must be equal to some constant $\lambda$. Dimensional analysis shows that this separation constant is the energy $E$ of the system.
 Therefore,
 \begin{eqnarray}
 \left[-\frac{\hbar^2}{2m}\frac{d ^2} {d x^2}+\frac{1}{2}m\omega^2 x^2 \right]\phi (x) = E\phi(x) \\
 \mbox{and} \hspace{0.5cm} i\hbar\frac{d}{dt}T(t) = E T(t)
 \end{eqnarray}
 Solving the above equation (Eq.(2.7)) in $T$ we get, 
 \begin{equation}
     T(t) = C e^{-\frac{iEt}{\hbar}}
 \end{equation}
  Now we wish to solve the differential equation for the space part. After rearranging, it can be written as  
 \begin{equation}
 \frac{d^2}{dx^2}\phi(x)-\frac{m^2 \omega^2}{\hbar^2}x^2\phi(x) = -\frac{2mE}{\hbar^2}\phi(x)
 \end{equation}
 which resembles with $\frac{d^2{y(x)}}{dx^2} + P(x)y(x) = Q y(x)$. The solution of Hermite differential equation can be written as a product of $e^{-cx^2} H_n (\alpha x)$. Here, $H_n(\alpha x)$  is the Hermite polynomial. To get the solution of this form we can use the series solution method of Frobenius. But here we want to try in a different way. We use Dirac's operator method to get the stationary state solution.
\section{Dirac's method}

 Here the differential equation contains a differential operator of second order. Dirac reduced this second order differential operator into two first order operators in factorized form. He defined two dimensionless operators $\hat{a}, \hat{a}^{\dag}$. These operators are
 \begin{eqnarray}
  \hat a = \frac{1}{\sqrt{2}}\left[\frac{\hat x}{\sqrt{\frac{\hbar}{m\omega}}} + \frac{i\hat{p}_x}{\sqrt{m\hbar\omega}}\right] \\
 \hat{a}^{\dag} = \frac{1}{\sqrt{2}}\left[\frac{\hat x}{\sqrt{\frac{\hbar}{m\omega}}} - \frac{i\hat{p}_x}{\sqrt{m\hbar\omega}}\right]
 \end{eqnarray}
 To make $\hat{a}$ and $\hat{a}^\dag$ dimensionless $\hat{x}$, $\hat{p}_x$ are divided by $\sqrt{\frac{\hbar}{m\omega}}$ and $\sqrt{m\hbar\omega}$ respectively. These two can also be written as 
 \begin{eqnarray}
 \hat a = \frac{1}{\sqrt{2m\hbar\omega}}[i\hat{p}_x + m\omega\hat x] \\
\hat {a}^{\dag} = \frac{1}{\sqrt{2m\hbar\omega}}[-i\hat{p}_x + m\omega\hat x]
 \end{eqnarray}
These two operators are called lowering and raising operators respectively. They are not hermitian operators  but their product is indeed a hermitian operator.
\subsection{Commutation relation of lowering and raising operator}
\begin{eqnarray}
[\hat{a}, \hat{a}^\dag] &=& \frac{1}{2m\hbar\omega}[i\hat{p}_x+m\omega\hat{x},-i\hat{p}_x+m\omega\hat{x}]\nonumber \\
&=& \frac{1}{2m\hbar\omega}([i\hat{p}_x+m\omega\hat{x},-i\hat{p}_x]
+[i\hat{p}_x+m\omega \hat{x},m\omega\hat{x}])\nonumber
\end{eqnarray}
\begin{eqnarray}
\hspace{0.85cm}=\frac{1}{2m\hbar\omega}([i\hat{p}_x,-i\hat{p}_x]+[m\omega\hat{x},-i\hat{p}_x]+ [i\hat{p}_x,m\omega\hat{x}] \nonumber\\
 + [m\omega\hat{x},m\omega\hat{x}]) \nonumber
 \end{eqnarray}
\begin{eqnarray}
&=& \frac{1}{2m\hbar\omega}(0-im\omega[\hat{x},\hat{p}_x]-im\omega[\hat{x},\hat{p}_x]+0)\nonumber\\
&=& \frac{1}{2m\hbar\omega}(-2im\omega \times i\hbar\hat{I})\nonumber \\
&=& \hat{I}
\end{eqnarray}
Therefore the required commutation relation is $[\hat{a}, \hat{a}^\dag] = \hat{I}$
\subsection{Position and momentum operators in terms of lowering and raising operators}
Adding Eq.(2.12) and (2.13) we get, 
\begin{eqnarray}
\hat{a}^\dag + \hat{a} = \frac{2m\omega}{\sqrt{2m\hbar\omega}}\hat{x} \nonumber \\
or, \hat{x} = \sqrt\frac{\hbar}{2m\omega}(\hat{a}^\dag + \hat{a}) 
\end{eqnarray}
Similarly on subtracting Eq.(2.12) from Eq.(2.13), we obtain 
\begin{equation}
\hat{p}_x = i\sqrt\frac{m\hbar\omega}{2}(\hat{a}^\dag - \hat{a})
\end{equation}
Now following these two equations we obtain \\ 
\begin{eqnarray}
\hat{x}^2 = \frac{\hbar}{2m\omega} (\hat{a}^\dag + \hat{a})^2 \nonumber \\ 
\mbox{and}\hspace{0.5cm} \hat{p}_x ^2 = -\frac{m\hbar\omega}{2} (\hat{a}^\dag - \hat{a})^2 \nonumber
\end{eqnarray}
\subsection{Hamiltonian of harmonic oscillator in terms of ladder operators}
To find Hamiltonian of harmonic oscillator in terms of ladder operators, first we find the product of ladder operators.
\begin{eqnarray}
\hat{a}^\dag\hat{a} &=& \frac{1}{2m\hbar\omega}(-i\hat{p}_x+m\omega \hat{x})(i\hat{p}_x+m\omega \hat{x})\nonumber\\
&=& \frac{1}{2m\hbar\omega}[(\hat{p}_x^2+m^2 \omega ^2 \hat{x}^2)-im\omega (\hat{p}_x\hat{x}-\hat{x}\hat{p}_x)]\nonumber \\
&=&\frac{1}{2m\hbar\omega}[(\hat{p}_x^2+m^2 \omega^2 \hat{x}^2)-im\omega [\hat{p}_x,\hat{x}]]\nonumber\\
&=& \frac{1}{2m\hbar\omega}[(\hat{p}_x^2+m^2 \omega ^2 \hat{x}^2)-im\omega (-i\hbar\hat{I})]\nonumber
\end{eqnarray}
Therefore,
\begin{align}
\hat{a^\dag}\hat{a} =& \frac{1}{2m\hbar\omega}(\hat{p}_x^2 + m^2 \omega ^2 \hat{x}^2)-\frac{1}{2}\hat{I} \nonumber\\
or, (\frac{\hat{p}_x^2}{2m} + \frac{1}{2}m \omega ^2 \hat{x}^2)=&(\hat{a}^\dag \hat{a} + \frac{1}{2}\hat{I})\hbar\omega \nonumber\\
or, \hat{H} =& (\hat{a}^{\dag}\hat{a} + \frac{1}{2}\hat{I})\hbar\omega
\end{align}
This is the Hamiltonian in terms of ladder operators. \\
Now we wish to find out the eigenvalues of this Hamiltonian. To do this, a new operator $\hat{N} = \hat{a}^{\dag}\hat{a}$ called number operator has to be introduced. It is named so because its eigen-values are natural numbers i.e. 0, 1, 2, 3, ..... .
Therefore, the Hamiltonian in terms of number operator can be written as 
\begin{equation}
\hat{H} = (\hat{N}+\frac{1}{2}\hat{I})\hbar\omega    
\end{equation}
Commutation relations of number operator with ladder operators are 
\begin{eqnarray}
[\hat{N}, \hat{a}] &=& [\hat{a}^\dag\hat{a},\hat{a}] \nonumber\\
&=& \hat{a}^\dag[\hat{a},\hat{a}]+[\hat{a}^\dag,\hat{a}]\hat{a}\nonumber \\
&=& \hat{a}^\dag.0+(-\hat{I})\hat{a} \nonumber\\
&=& -\hat{a} 
\end{eqnarray}
and
\begin{eqnarray}
[\hat{N},\hat{a}^\dag] &=& [\hat{a}^\dag\hat{a},\hat{a}^\dag] \nonumber\\
&=&\hat{a}^\dag[\hat{a},\hat{a}^\dag] + [\hat{a}^\dag,\hat{a}^\dag]\hat{a} \nonumber\\
&=&\hat{a}^\dag\hat{I} + 0.\hat{a} \nonumber\\
&=&\hat{a}^\dag 
\end{eqnarray}
Meanwhile, the commutation relation of the number operator $\hat{N}$ and the Hamiltonian $\hat{H}$ is
\begin{equation}
[\hat{N},\hat{H}]=[\hat{a}^\dag\hat{a},\hat{a}^{\dag}\hat{a} + \frac{1}{2}\hat{I}]\hbar\omega = 0  
\end{equation}
As $\hat{H}$ commutes with $\hat{N}$ then both have complete set of common eigen states.
\subsection{Eigen values and eigen functions of number operator}
\vspace{0.3cm}
Let us consider $\vert \phi_{\lambda}\rangle$ as the eigen state and $\lambda$ is the corresponding eigen value of number operator $\hat{N}$. Here, the eigen value $\lambda$ must be a real number, as the number operator is a hermitian operator. Thus, 
\begin{equation}
 \hat{N}\vert \phi_{\lambda}\rangle = \lambda \vert \phi_{\lambda} \rangle    
\end{equation}
Now using operator equation $[\hat{N}, \hat{a}] = -\hat{a}$ (Eq.(2.19)), we obtain
\begin{eqnarray}
[\hat{N}, \hat{a}]\vert \phi_{\lambda}\rangle &=& -\hat{a} \vert \phi_{\lambda}\rangle  \nonumber\\
or, \hat{N}\hat{a}\vert\phi_\lambda\rangle-\hat{a}\hat{N} \vert \phi_\lambda\rangle &=& -\hat{a} \vert\phi_\lambda\rangle \nonumber \\
or,\hat{N}\hat{a}\vert\phi_\lambda\rangle-\lambda\hat{a}\vert \phi_\lambda\rangle &=& -\hat{a} \vert \phi_\lambda\rangle \nonumber \\
or, \hat{N}(\hat{a} \vert \phi_\lambda\rangle) = (\lambda - 1)(\hat{a} \vert \phi_\lambda\rangle)
\end{eqnarray}
This implies that if $\vert \phi_\lambda\rangle$ is an eigen state of the number operator $\hat{N}$, then $\hat{a} \vert \phi_{\lambda}\rangle$ also be an eigen state of $\hat{N}$, but its corresponding  eigen value is reduced by one unit.\\
Similarly from equation (2.20), $[\hat{N},\hat{a}^{\dag}] = \hat{a}^{\dag}$ we get
\begin{eqnarray}
[\hat{N},\hat{a}^{\dag}]\vert\phi_{\lambda}\rangle &=& \hat{a}^{\dag} \vert\phi_{\lambda} \rangle \nonumber\\
or, \hat{N}(\hat{a}^{\dag}|\phi_\lambda\rangle) &=& (\lambda + 1)\hat{a}^{\dag}\vert\phi_{\lambda}\rangle
\end{eqnarray}

It says that if $\vert \phi_\lambda\rangle$ is an eigen state of $\hat{N}$, then $\hat{a}^{\dag}\vert \phi_{\lambda}\rangle$ also is an eigen state of $\hat{N}$ with eigen value $(\lambda + 1)$ which is raised by one unit. That is why the operators $\hat{a}^{\dag}$ and $\hat{a}$ are called raising and lowering operator respectively. 

Now if we start with $\vert \phi_\lambda\rangle$ and proceed in the same way as 
\begin{eqnarray}
[\hat{N},\hat{a}^2]&=&[\hat{a}^\dag\hat{a},\hat{a}\hat{a}]\nonumber\\
&=&\hat{a}^\dag[\hat{a},\hat{a}\hat{a}]+[\hat{a}^\dag,\hat{a}\hat{a}]\hat{a}\nonumber\\
&=& 0+\left([\hat{a}^{\dag},\hat{a}]\hat{a}\hat{a}+\hat{a}[\hat{a}^{\dag},\hat{a}]\hat{a}\right)\nonumber\\
&=&-2\hat{I}\hat{a}^2\nonumber\\
&=&-2\hat{a}^2\nonumber
\end{eqnarray}
So, $[\hat{N},\hat{a}^2] = -2\hat{a}^2$.\\
Now we see that $\hat{a}^2 \vert \phi_\lambda\rangle$ is also an eigen state of $\hat{N}$. Using the above relation we get
\begin{eqnarray}
[\hat{N}, \hat{a}^2] \vert \phi_\lambda\rangle &=& -2\hat{a}^2 \vert\phi_\lambda\rangle \nonumber\\
or, \hat{N}\hat{a}^2 \vert \phi_\lambda\rangle-\hat{a}^2\hat{N} \vert \phi_\lambda\rangle &=&-2\hat{a}^2|\phi_\lambda\rangle \nonumber\\
or, \hat{N}\hat{a}^2 \vert \phi_\lambda\rangle - \lambda(\hat{a}^2 \vert \phi_\lambda\rangle) &=& -2\hat{a}^2 \vert \phi_\lambda\rangle\nonumber\\
or, \hat{N}\hat{a}^2 \vert \phi_\lambda\rangle &=& (\lambda -2)\hat{a}^2 \vert \phi_\lambda\rangle\nonumber
\end{eqnarray}
Therefore, $\hat{a}^2\vert \phi_\lambda\rangle$ is also an eigen state of $\hat{N}$ with eigen value $(\lambda-2)$. If we continue operating using $\hat{a}^3$, $\hat{a}^4$, $\hat{a}^5$, .... and so on successively, at some point we may get an eigen state with negative eigen value. But it is not possible. We can't have a negative eigen value for $\hat{N}$. Actually it has a lower bound (i.e. zero) and no upper bound. In the following section we will show the reason and also present some properties of eigen value $\lambda$.
\subsection{ Properties of eigen values of number operator}
\subsubsection{\textit{Lemma-1:} Eigen values of $\hat{N}$ are either real positive or zero}
To prove this Lemma we follow the book Quantum Mechanics by C. Cohen-Tannoudji et.al\cite{Cohen}. First we consider that $\hat{N}$ has an arbitrary, non zero normalized eigen state $\vert \phi_\lambda\rangle$ and its corresponding  eigen value $\lambda$.
\begin{eqnarray}
\hat{N} \vert \phi_{\lambda}\rangle &=& \lambda \vert \phi_{\lambda}\rangle,  \lambda \mbox{ is real} \nonumber\\
\langle\phi_\lambda\vert \hat{N} \vert \phi_\lambda\rangle &=& \lambda \langle\phi_\lambda\vert \phi_\lambda\rangle \nonumber \\
or, \langle\phi_\lambda|\hat{a}^\dag\hat{a}\vert \phi_\lambda\rangle &=& \lambda [\mbox{ as,  } \langle\phi_\lambda|\phi_\lambda\rangle = 1] \nonumber\\
or, \langle\hat{a}\phi_\lambda \vert \hat{a}\phi_\lambda\rangle &=& \lambda \nonumber \\ 
or, \vert \vert \hat{a}|\phi_\lambda\rangle \vert \vert ^2 &=&\lambda
\end{eqnarray}
It is seen that $\vert \vert \hat{a}\vert\phi_\lambda\rangle \vert \vert^2 = 0$ implies  $\lambda = 0$. Otherwise $\vert \vert \hat{a} \vert \phi_\lambda\rangle \vert \vert ^2 > 0$ implies $\lambda > 0$. So, $\lambda \geqslant 0$. Moreover the number operator is hermitian then the eigen value $\lambda$ must be real. Up to now we have seen that  $\lambda$'s are real positive or zero.
\subsubsection{\textit{Lemma-2:} Eigen values of $\hat{N}$ are non-negative integers}
In the \textit{\textbf{lemma-1}}, we have seen that the eigen values of number operator are either non-zero real positive or zero. Now, it can be shown that $\lambda$'s can take only positive integer values including zero[9]. To do this, first we assume that $\lambda$ to be non-integral. We are now going to show that such a hypothesis contradicts the \textit{\textbf{lemma-1}}.\\
If $\lambda$ is non integral, then we can always find an integer $n\geqslant 0$ such that $n < \lambda <n+1 $.
Now let us consider the series $\vert \phi_{\lambda}\rangle$, $\hat{a} \vert \phi_{\lambda}\rangle$, $\hat{a}^2\vert \phi_{\lambda}\rangle$, ...., $\hat{a}^{n} \vert \phi_{\lambda}\rangle$. According to our assumption, each of the vector $\hat{a}^{p} \vert \phi_{\lambda}\rangle$ of this series (for $0 \leqslant p \leqslant n$) is non zero eigen vectors of $\hat{N}$ with eigen value $(\lambda - p)$.\\
In the following table eigen vectors and their corresponding eigen values are shown.\\
\begin{table}[h]
\centering
\begin{tabular}{|c|c|}
\hline
\textbf{Eigen vector} &\textbf{Eigen value}\\
\hline
$\vert \phi_{\lambda}\rangle$ & $\lambda$ \\ 
\hline
$\hat{a}\vert\phi_{\lambda}\rangle$  &$\lambda-1$\\
\hline
..... & .....\\
\hline
$\hat{a}^{n-1}\vert \phi_{\lambda}\rangle$ & $\lambda-n+1$\\
\hline
$\hat{a}^{n}\vert\phi_{\lambda}\rangle$ &  $\lambda-n$ \\ 
\hline
$\hat{a}^{n+1}\vert\phi_{\lambda}\rangle$ & $\lambda-n-1$ \\ 
\hline
\end{tabular}
\caption{ Eigen-states of number operator and their corresponding eigen values are tabulated here.}
\label{table:1}
\end{table}
\textit{\textbf{ Proof of the Lemma-2 using iteration:}}\\\\
Consider $\vert\phi_\lambda\rangle $ is a non zero vector; $\hat{a}\vert\phi_\lambda\rangle$ is also an non zero eigen vector of $\hat{N}$ as $\lambda > 0 $ and it corresponds to eigen value $(\lambda-1)$. Now we successively apply $\hat{a}$ and reach to a state $\hat{a}^n\vert \phi_\lambda\rangle$; it is also an eigen state of $\hat{N}$ with an eigen value $(\lambda-n)$. Similarly, we say that $\hat{a}^{n+1}\vert \phi_\lambda\rangle$ is also an eigen state of $\hat{N}$ with eigen value $(\lambda-n-1)$  which is strictly negative according to $n <\lambda <(n+1)$. If $\lambda$ is non-integral, one can construct a non zero eigen vector of $\hat{N}$ with a strictly negative eigen value. This fact is impossible according to \textit{\textbf{lemma-1}}. So, the hypothesis about non integral $\lambda$ value must be rejected.\\
What happens if $\lambda = n$ where, $n$ takes the value of positive integer or zero? In the series of eigen vectors shown in the table above, $\hat{a}^n\vert \phi_\lambda\rangle$ is non zero  eigen vector of $\hat{N}$ with $0$ eigen value i.e. 
\begin{equation}
 \hat{N}(\hat{a}^n\vert\phi_\lambda\rangle) = 0.(\hat{a}^n\vert\phi_\lambda\rangle) \nonumber  
\end{equation}
Let us consider the state $(\hat{a}^n\vert \phi_\lambda\rangle)$ be denoted as $\vert \phi_0\rangle$. 
Therefore,
\begin{eqnarray}
\hat{N}\vert \phi_0\rangle &=& 0 \vert \phi_0\rangle\nonumber \\
or, \hat{a}^\dag \hat{a}\vert \phi_0\rangle &=& 0 \vert \phi_0\rangle\nonumber\\
or, \langle\phi_0\vert \hat{a}^\dag \hat{a}\vert \phi_0\rangle &=& 0  \langle\phi_0\vert \phi_0\rangle \nonumber \\
or, \vert \vert \hat{a}\vert \phi_0\rangle\vert \vert^2 &=& 0 [\mbox{as,  } \langle\phi_0\vert\phi_0\rangle = 1] \nonumber
\end{eqnarray}
\begin{eqnarray}
\therefore \hat{a}|\phi_0\rangle = \Vec{0} \nonumber
\end{eqnarray}
It means that further application of annihilation operator $\hat{a}$ annihilates the state $\vert \phi_0\rangle$.\\
Finally, we claim that $\lambda$ belongs to the positive integer set including $0$, or $\lambda = n$; $n = 0,1,2,3,...$
In traditional notation we can write $\hat{N}\vert n\rangle = n\vert n\rangle$; $n = 0,1,2,3,...$ 
\subsection{Number states ${\vert n\rangle}$form a complete basis set}
\vspace{0.3cm}
Let, $ f \epsilon L^2(C)$; $L^2$ is a space of square summable sequences.
\begin{equation}
   \vert f\rangle = \sum_{m=0} ^{\infty}a_m\vert m\rangle 
\end{equation}
Let's act $\sum_{n=0}^{\infty}\vert n\rangle \langle n\vert$ from left on both sides of the above equation.
\begin{eqnarray}
\left(\sum_{n=0}^{\infty}\vert n\rangle \langle n\vert\right) \vert f\rangle 
&=&\sum_{n=0}^{\infty}\vert n\rangle \langle n|.\sum_{m=0} ^{\infty}a_m\vert m\rangle \nonumber\\
&=&\sum_{n=0} ^{\infty}\sum_{m=0}^{\infty}a_m \vert n\rangle\langle n\vert m\rangle\nonumber\\ 
&=& \sum_{m=0}^{\infty}a_m\vert m\rangle  [\mbox{as  } \langle n\vert m \rangle =\delta_{nm}] \nonumber\\ 
&=&\vert f\rangle\nonumber
\end{eqnarray}
So, we find that $\sum_{n = 0} ^{\infty}\vert n\rangle\langle n\vert = \hat{I}$. It is the completeness relation in number state basis.
\subsection{Energy eigen values of harmonic oscillator}
Now we have the Hamiltonian of harmonic oscillator in terms of number operator
\begin{equation}
   \hat{H} = (\hat{N}+\frac{1}{2}\hat{I})\hbar\omega \nonumber 
\end{equation}
Therefore,
\begin{eqnarray}
\hat{H}\vert n\rangle &=& (\hat{N}+\frac{1}{2}\hat{I})\hbar\omega\vert n\rangle\nonumber\\
or, \hat{H}\vert n\rangle &=& (n+\frac{1}{2})\hbar\omega\vert n\rangle\nonumber\\
or, \hat{H}\vert n\rangle &=& E_n\vert n\rangle \nonumber
\end{eqnarray}
$\therefore E_n =(n+\frac{1}{2})\hbar\omega $ is the energy eigen value corresponding to the state $\vert n \rangle$.\\
Now $n = 0$ corresponds to ground state energy $E_0 = \frac{1}{2}\hbar\omega$; the minimum value. It is called zero point energy. So the lowest possible energy is not zero but $\frac{1}{2}\hbar\omega$. It appears due to quantum fluctuations. It essentially says that in harmonic potential $[V(x) = \frac{1}{2}kx^{2}]$; the classically equilibrium state is that the particle sitting at the equilibrium point (at origin $(x = 0)$) at rest means zero kinetic energy and zero potential energy. But in quantum mechanics it is not possible because if we try to localize the particle at $x=0$; its momentum becomes infinity. So we can't have a state with absolute rest.
\subsection{Excited state}
\vspace{0.2cm}
Traditionally we designate $\vert n\rangle$ as $n$-th excited state of harmonic oscillator. $\vert 0\rangle$, $\vert 1\rangle$, .... represents ground state, first excited state, ... of harmonic oscillator. We have $\vert 1\rangle = \hat{a}^{\dag}\vert 0\rangle$; $\vert 2\rangle = \hat{a}^{\dag}\vert 1\rangle$. Now in general 
\begin{equation}
  \vert n+1 \rangle = \hat{a}^{\dag}\vert n\rangle
\end{equation}
Here we consider that $\vert n\rangle$ is normalized but $\vert n+1\rangle$ is not normalized. So, we need to normalize it. Now we multiply $\hat{a}^{\dag}\vert n\rangle $ with some constant $C_n$ which makes $\vert n+1\rangle $ normalized.
\begin{eqnarray}
\vert n+1 \rangle&=& C_n\hat{a}^{\dag}\vert n\rangle \nonumber\\
or, \langle n+1\vert n+1\rangle&=&\vert C_n\vert ^2\langle n\vert \hat{a}\hat{a}^{\dag}\vert n\rangle\nonumber\\
or, 1&=& \vert C_n\vert^2\langle n\vert (\hat{N}+\hat{I})\vert n\rangle [\mbox{as,} [\hat{a},\hat{a}^{\dag}] = \hat{I}
]\nonumber\\
or, 1 &=& \vert C_n\vert ^2(n+1)\nonumber\\
or, C_n &=& \frac{1}{\sqrt{n+1}}e^{i\alpha}\nonumber.
\end{eqnarray}
As measurement in quantum mechanics is probabilistic the phase $\alpha$ is irrelevant.
Therefore, $C_n=\frac{1}{\sqrt{n+1}}$
Now, we can write,
\begin{eqnarray}
\vert n+1\rangle &=& \frac{1}{\sqrt{n+1}}\hat{a}^{\dag}\vert n\rangle \nonumber\\
or, \hat{a}^{\dag}\vert n\rangle &=& \sqrt{n+1}\vert n+1\rangle; [n=0,1,2,.]
\end{eqnarray}
In a similar way we can write,
\begin{equation}
 \vert n-1\rangle = d_n\hat{a}\vert n\rangle ;n\neq 0\nonumber   
\end{equation}
 where $d_n$ is the normalization constant.
\begin{eqnarray}
\langle n-1\vert n-1\rangle &=& \vert d_n\vert ^2\langle n|\hat{a}^{\dag}\hat{a}|n\rangle\nonumber\\
or,1 &=& \vert d_n\vert ^2\langle n\vert \hat{N}\vert n\rangle\nonumber\\
or, 1 &=& \vert d_n\vert^{2}n\nonumber\\
or, d_n &=& \frac{1}{\sqrt{n}}\nonumber
\end{eqnarray}
 \hspace{3cm}\\
Thus, ignoring phase term 
\begin{eqnarray}
\vert n-1 \rangle &=& \frac{1}{\sqrt{n}}\hat{a} \vert n\rangle\nonumber\\
or, \hat{a}\vert n\rangle &=& \sqrt{n} \vert n-1\rangle; n=1,2,...
\end{eqnarray}
Now for $n = 1$, $\hat{a} \vert 1\rangle = \vert 0\rangle$.
Earlier we have shown that $\hat{a}\vert 0\rangle = \vec{0}$(zero vector).
\subsection{Wave functions in position basis}
\subsubsection{Ground state wave function:}
\vspace{0.3cm}
Now we wish to find out the form of ground state wave function in position basis. For ground state we have, 
\begin{eqnarray}
\hat{a} \vert 0 \rangle &=& \vec{0} \nonumber\\
or, \frac{1}{\sqrt{2m\hbar\omega}}(i\hat{p_x}+m\omega\hat{x})\vert 0\rangle &=&\vec{0}\nonumber\\
or, (i\hat{p_x}+m\omega\hat{x})\vert 0\rangle &=&\vec{0} \nonumber \\
or, \langle x\vert i\hat{p_x} + m\omega\hat{x}) \vert 0\rangle &=& 0\nonumber
\end{eqnarray}
 where ${\vert x\rangle}$ represents position basis.
 \begin{eqnarray}
 \hbar\frac{d}{dx}\langle x \vert 0\rangle + m\omega x\langle x\vert 0\rangle &=& 0\nonumber \\
or, \hbar\frac{d}{dx}\phi_0(x) + m\omega x \phi_0(x) &=& 0 \nonumber
 \end{eqnarray}
where, $\langle x\vert 0\rangle = \phi_0(x).$

Therefore,
\begin{equation}
\frac{d\phi_0}{dx} + \frac{m\omega x}{\hbar}\phi_0(x) = 0
\end{equation}
Solving the above equation we get 
\begin{equation}
   \phi_0(x) = A_{0} e^{{-\frac{m\omega x^{2}}{2\hbar}}} 
\end{equation}
where, $A_{0}$ is the normalization constant.
The ground state wave function is a Gaussian function with no nodes. Now the normalization condition
$\int_{-\infty}^{+\infty}\vert \vert \phi_0(x)\vert \vert^{2} dx = 1$ 
gives
$A_{0} = (\frac{m\omega}{\pi\hbar})^\frac{1}{4}$.
Therefore, the ground state wave function is
\begin{equation}
    \phi_0(x) = \left(\frac{m\omega}{\pi\hbar}\right)^\frac{1}{4}e^{-\frac{m\omega}{2\hbar}x^{2}}
\end{equation}
\subsubsection{First excited state wave function:}
First excited state wave function can be found out easily as follows,
\begin{eqnarray}
\vert 1\rangle &=& \hat{a}^{\dag}\vert 0\rangle\nonumber \\
or, \langle x\vert 1\rangle &=& \langle x\vert \hat{a}^{\dag} \vert 0\rangle \nonumber\\ 
or, \phi_1(x) &=& \langle x \vert \hat{a}^{\dag}\vert 0\rangle  \mbox{ where, } \langle x\vert 1\rangle=\phi_1(x) \nonumber \\ 
or, \phi_1(x) &=& \langle x\vert\frac{1}{\sqrt{2m\hbar\omega}}(-i\hat{p_x} + m\omega \hat{x})\vert 0\rangle \nonumber \\
or, \phi_1(x) &=& \frac{1}{\sqrt{2m\hbar\omega}}(-\hbar\frac{d}{dx}+m\omega x)\langle x \vert 0\rangle \nonumber\\
or, \phi_1(x) &=& \frac{1}{\sqrt{2m\hbar\omega}}(-\hbar\frac{d}{dx} + m\omega x)\phi_0 (x)) \nonumber \\
or, \phi_1(x) &=& \frac{1}{\sqrt{2m\hbar\omega}}[-\hbar\frac{d\phi_0(x)}{dx} + m\omega x\phi_0 (x)] \nonumber \\
or, \phi_1(x) &=& \sqrt{\frac{m\omega}{2\hbar}}x\phi_0(x)-\sqrt{\frac{\hbar}{2m\omega}}\frac{d\phi_0(x)}{dx} 
\end{eqnarray}
Since, $\phi_0(x) = (\frac{m\omega}{\pi\hbar})^\frac{1}{4}e^{-\frac{m\omega }{2\hbar}x^2}$ we can easily obtain $\phi_1(x)$ by substituting $\phi_0(x)$ in the above equation. It looks like
\begin{equation}
 \phi_{1}(x) = A_{1} x e^{-\frac{m\omega}{2\hbar}x^2}   
\end{equation}
where, $A_{1}$ is some normalization constant, can be found out using the normalization condition. Since, $\phi_0(x)$ is an even function; then $\phi_{1}(x)$ must be an odd function.  All the eigen functions of harmonic oscillator Hamiltonian must have definite parities. This is so, because the Hamiltonian of harmonic oscillator commutes with parity operator $\hat\Pi$.\\
The higher order wave functions can be found out in similar way and they have the form 
\begin{equation}
 \phi_{n} (x) = A_{n} H_{n} \left(x\sqrt{\frac{m\omega}{\hbar}}\right)e^{-\frac{m\omega}{2\hbar}x^{2}}   
\end{equation}
where, $A_{n}$ is the normalization constant for $n$-th order wave function; $H_{n} \left(x\sqrt{\frac{m\omega}{\hbar}}\right)$ is the Hermite polynomial which actually carries the parity of wave function.
\section{Correspondence between Linear harmonic oscillator and Quantized electromagnetic field of radiation}
\vspace{0.3cm}
Why we discuss linear harmonic oscillator with such an emphasis? There is one to one correspondence between the linear harmonic oscillator problem and quantum optics i.e. the framework in which we discuss quantized electromagnetic field is the same as linear harmonic oscillator. In case of quantum optics we try to understand the various states of the quantized electromagnetic field of radiation.\\
There is an algebraic equivalence between linear harmonic oscillator and quantized beam of light. The raising and lowering operators $\hat{a},  \hat{a}^\dag$ are correspondingly equivalent to the annihilation and creation operator in case of quantized electromagnetic field of radiation. $n$ is a label for the energy eigen state for harmonic oscillator whereas in quantized electromagnetic radiation field it is the number of photons in a given state. $\vert 0\rangle$,  $\vert 1\rangle$ are said to be as ground state and first excited state of harmonic oscillator but in other cases  $\vert 0\rangle$, $\vert 1\rangle$ are called zero photon or vacuum state and one photon state respectively.\\
If we operate $\hat{a}^{\dag}$ for $n$ times on an energy eigen state; it goes to higher excited state of the oscillator but in case of quantized field  it is going to add energy $n\hbar\omega$; that means it creates $n$ number of photons. So in the later case $n$ is no longer a label, it refers to $n$ number of photons. $\hat{a}^{\dag}\hat{a}$ is now described as photon number operator. It counts the number of photons in the state.
\section{Algebraic equivalence between single mode quantized beam of light and one dimensional linear harmonic oscillator}
\vspace{0.3cm}
To show the equivalence we need a field-theoretic approach for the single mode light beam in which electric field $E$ and magnetic field $B$ are considered as two non commuting field operators. Now we talk about photon states. Photon states are the quantum states of the electromagnetic field. Discrete piece of energy and momentum are carried by this particle. When we talk about photon states we really need quantum field theory (Photon states have a connection with electromagnetic field-electric and magnetic field. One important quantity about electromagnetic field is its energy. We know that energy is Hamiltonian in classical sense). Now we consider $\Vec{k}$ is the propagation vector of the single-mode (single frequency/ wavelength)field. This is a field which is consistent with Maxwell’s equations with some boundary conditions. Direction of $\Vec{k}$ is along z-direction. This field is in a cavity of length $L$ which extends along Z-direction. Electric field $\vec{E}$ is given by
\begin{equation}
  \vec{E} = \vec{e}_x \sqrt{\frac{2\omega^2}{\epsilon_0 V}} q(t) sin kz  
\end{equation}
where $\vec{e}_x$ is the unit vector in the direction of polarization, $V$ volume of the cavity, $\epsilon_0$ permittivity of free space, frequency of wave $\omega = ck$, $q(t)$ some function of time as electromagnetic field evolves in space time. Here, $sin kz$  is due to boundary condition $\vec{E}(z=0)=0,\vec{E}(z=L)=0$ gives $sin kL= 0 = sin m\pi; k = \frac{m\pi}{L}; m=1,2,3,...$. Now, $\omega= \frac{m\pi}{L}c$.
For simplicity we consider $\omega =  \frac{\pi}{L}c$. This problem of quantized electromagnetic field propagating along z-direction in a cavity has a one-to one correspondence with linear harmonic oscillator. Generally, it is seen that the field, (as in Eq.(2.36)) 
$\vec{E} = \vec{e}_x \sqrt{\frac{2\omega^2}{\epsilon_0 V}} q(t) sin kz$,
satisfies the free-field Maxwell's equations
\begin{subequations}
\begin{align}
\nabla \cdot \vec{E} &= 0\label{eqn:line-1} \\
\nabla \cdot \vec{B} &= 0 \label{eqn:line-2} \\
\nabla\times\vec{E} &= -\frac{\partial \vec{B}}{\partial t} \label{eqn:line-3} \\
\nabla\times\vec{B} &= \frac{1}{c^2}\frac{\partial \vec{E}}{\partial t}
\end{align}
\label{eqn:all-lines}
\end{subequations}
Here, we assume that there are no free charges or currents inside the cavity. For the given electric field one can find the magnetic field
\begin{eqnarray}
\vec{B} &= \hat{e}_y\frac{1}{c}\sqrt{\frac{2}{\epsilon_0 V}}\Dot{q} cos kz \nonumber\\ 
or, \vec{B} &=\hat{e}_y\frac{1}{c}\sqrt{\frac{2}{\epsilon_0 V}}p(t) cos kz 
\end{eqnarray}
where, $p(t)=\Dot{q}$\\
We are now equipped with the electromagnetic field configuration. Hamiltonian for the electromagnetic field is given by
\begin{equation}
  H = \frac{1}{2}\int dV \epsilon_0(E^2 + c^2 B^2) 
\end{equation}
 The bracketed term multiplied by $\epsilon_0$ inside the integral is called energy density which is integrated over the entire volume gives the total energy. Now we calculate the integral in two steps. \\
First Part of the integral:
\begin{eqnarray}
\frac{1}{2}\epsilon_0\int{dx dy dz}E^2&=&\frac{1}{2}\epsilon_0\frac{2\omega ^2 q^2(t)}{\epsilon_0 V}\int_0^L dz sin^2 kz\int\int dx dy \nonumber\\
&=&{\frac{\omega^2 q^2(t)}{V}}{\int_0^L dz \frac{1-cos2kz}{2}\int\int dx dy} \nonumber \\
&=& \frac{1}{2}\omega^2 q^2(t)
\end{eqnarray}
Second part of the integral:
 \begin{eqnarray}
\frac{1}{2}\epsilon_0 c^2 \int dx dy dz B^2 
&=& \frac{1}{2} \epsilon_0 c^2\frac{2p^2(t)}{c^2\epsilon_0 V}\int_0^L dz cos^2kz\int\int dx dy \nonumber\\
&=& \frac{p^2(t)}{V}\int_0^L dz \frac{(1+cos2kz)}{2}\int\int dxdy \nonumber\\
&=& \frac{1}{2}p^2(t)
 \end{eqnarray}
Hamiltonian for the electromagnetic field becomes
\begin{equation}
 H = \frac{1}{2}(\omega^2 q^2(t)+p^2(t))   
\end{equation}
This Hamiltonian looks like the Hamiltonian for Linear harmonic oscillator except the fact that mass $m$ is absent in the field Hamiltonian. To describe electromagnetic field in quantum mechanical way we need to promote $q(t)$, $p(t)$ as Heisenberg operators. Thus the Hamiltonian of the quantized single mode radiation field is
\begin{equation}
   \hat{H} = \frac{1}{2}(\omega^2\hat{q}^2(t)+\hat{p}^2(t)) 
\end{equation}
Now it is more convenient to deal with a set of non hermitian operators $\hat{a}$, $\hat{ a}^{\dag}$ rather than hermitian operators $\hat {q}$, $\hat{p}$. The non hermitian operators are defined as
\begin{eqnarray}
\hat{a}(t) = \frac{1}{\sqrt{2\hbar\omega}} [\omega \hat{q}(t) + i\hat{p}(t)] \\
\hat{a}^{\dag}(t) = \frac{1}{\sqrt {2\hbar\omega}}[\omega\hat{q}(t) - i\hat{p}(t)]
\end{eqnarray}
Now we can write $\hat{q}(t)$, $\hat{p}(t)$ in terms of $\hat{a}(t)$, $\hat{a}^{\dag}(t)$ as 
\begin{eqnarray}
\hat{q}(t) = \sqrt{\frac{\hbar}{2\omega}}[\hat{a}(t)+\hat{a}^{\dag}(t)]\\  \hat{p}(t) = i\sqrt{\frac{\hbar\omega}{2}}[\hat{a}^{\dag}(t)-\hat{a}(t)]
\end{eqnarray}

The Hamiltonian
       \[\hat{H}(t) = \frac{1}{2}(\omega^2\hat{q}^2(t)+\hat{p}^2(t))\]

 can also be expressed in terms of annihilation and creation operator i.e.
 \begin{equation}
  \hat{H}(t) = [\hat{a}^{\dag}(t)\hat{a}(t)+\frac{1}{2}\hat{I}]\hbar\omega  
 \end{equation}
 
 We can write $\hat{H}$ in a more compact form
\begin{equation}
     \hat{H}=[\hat {N}+\frac{1}{2}\hat{I}]\hbar\omega 
\end{equation}
 where, $\hat{N}$ is the photon occupation number operator. Its eigen value equation is
 \begin{equation}
     \hat{N} \vert n\rangle = n\vert n\rangle 
 \end{equation}
$n = 0,1,2,3,....$ and $\vert n\rangle$ is the photon occupation state.\\
The Hamiltonian in Eq.(2.49) looks similar to the Hamiltonian for Linear harmonic oscillator apart from the phase factor.\\

 Now we find the time dependence of the operators $\hat{a}(t)$, $\hat{a}^{\dag}(t)$ solving Heisenberg's equation of motion.
 \begin{equation}
 \frac{d\hat{a}}{dt} =
 \frac{i}{\hbar}\left[\hat{H},\hat{a}\right] = -i\omega\hat{a}\nonumber
 \end{equation}
 which has the solution $\hat a (t) = e^{-i\omega t}\hat a(0)$. In a similar way $\hat a ^{\dag}(t) = e^{i\omega t}\hat{a}^{\dag}(0)$.

 Using the solutions above we write
 \begin{equation}
  \hat{q}(t)=\sqrt{\frac{\hbar}{2\omega}}[e^{-i\omega t}\hat{a}(0)+e^{i\omega t}\hat{a}^{\dag}(0)]   
 \end{equation}
 We substitute $\hat{q}(t)$  in the Eq.(2.36) 
  \[\vec{\hat{E}} = \vec{e}_x \sqrt{\frac{2\omega^2}{\epsilon_0 V}} \hat{q}(t) sin kz\]  
 and find that 
  \begin{equation}
       \vec{\hat{E}} (z,t) = \hat{e}_x \sqrt{\frac{\hbar\omega}{\epsilon_0 V}}[e^{-i\omega t}\hat{a}(0)+e^{i\omega t}\hat{a}^{\dag}(0)]sin kz
  \end{equation}
 It is seen that $\vec{\hat{E}} (z,t)$ is hermitian. 

Now we try to find the expectation value of $\hat{E} (z,t)$ with respect to a photon occupation state $\vert n\rangle$.

Expectation value:
\begin{eqnarray}
\langle n \vert \vec{\hat{E}} \vert n\rangle = 
\hat{e}_x\sqrt{\frac{\hbar\omega}{\epsilon_0 V}}[e^{-i\omega t}\langle n \vert \hat{a}(0)\vert n\rangle + \nonumber\\
e^{i\omega t}\langle n \vert \hat{a}^{\dag}(0)\vert n\rangle]sinkz = 0
\end{eqnarray}
Actually this is not too strange. $\vert n\rangle$ is the state in which photon occupation does not change with time. At this point we introduce a different  state, coherent state $\vert z\rangle$ in this context. This state is as good as classical state. It is also an eigen state of photon annihilation operator $\hat{a}$. Coherent state is produced by the action of displacement operator on the vacuum state i.e. 
\begin{eqnarray}
 \vert z\rangle &=& exp(z \hat{a}^{\dag}-z^*\hat{a})\vert 0\rangle \nonumber\\
&=& exp\left(-\frac{\vert z \vert^2}{2}\right)\sum_{n=0}^{\infty} \frac{z^n}{\sqrt{n!}}\vert n\rangle \nonumber\\
\end{eqnarray}
and,
\begin{equation}
    \hat{a}\vert z\rangle = z\vert z\rangle  
\end{equation}
Coherent state can accommodate infinite number of photons. If we take the expectation value of $\hat{\vec{E}}$ with respect to coherent state $\vert z\rangle$ we find  
\begin{eqnarray}
\langle z\vert \hat{\vec{E}} \vert z\rangle 
&=& \hat{e}_x\sqrt{\frac{\hbar\omega}{\epsilon_0 V}}[e^{-i\omega t}\langle z\vert \hat{a}(0)\vert z \rangle + e^{i\omega t}\langle z\vert \hat{a}^{\dag}(0)\vert z\rangle]sin kz \nonumber\\
&=& \hat{e}_x\sqrt{\frac{\hbar\omega}{\epsilon_0 V}}[e^{-i\omega t}z+e^{i\omega t}z^{*}]sin kz\nonumber\\
&=& 2\hat{e}_x\sqrt{\frac{\hbar\omega}{\epsilon_0 V}}Re(ze^{-i\omega t})sin kz\nonumber\\
&=& 2\hat{e}_x\sqrt{\frac{\hbar\omega}{\epsilon_0 V}}|z|cos(\omega t-\theta)sin kz
\end{eqnarray}
This is something like a standing wave that changes in time with a fixed spatial distribution. It is a good description classically as well as quantum mechanically. The above expression implies that coherent state is the state with respect to which the expectation value of $\hat{\vec{E}}$ is precisely the kind of wave seems to be classical. This wave appears because coherent state is the right and left eigen state of $\hat{a}, \hat{a}^\dag$ respectively. For this reason $\hat{a},\hat{a}^\dag$  has expectation values with respect to coherent state $\vert z\rangle$ . We call– a classical wave resonating in a cavity corresponds to the coherent state of the electromagnetic field in quantum mechanical sense. We observe that the funny superposition of several photon occupation states form a coherent state. It is not an energy eigen state; not an eigen state of photon occupation operator$\hat{N}$ . This state possesses nice classical picture. Expectation value of $\hat{\vec{E}}$ with respect to  $\vert z\rangle$  is real. When we try to analyze classical wave; quantum description of that wave is the coherent state of the electromagnetic field.\\  
Now we try to show that the single mode quantized beam of light corresponds to a linear harmonic oscillator of that particular mode. This connection is purely mathematical but logical. All we know that coherent state $\vert z\rangle$ is the eigen state of photon annihilation operator. Suppose its time evolution is governed by time dependent Schrodinger equation i.e.
\begin{equation}
 i\hbar\frac{d\vert z(t)\rangle }{dt} = \hat{H}\vert z(t)\rangle   
\end{equation}
Here we consider Hamiltonian is time-independent.
\begin{eqnarray}
\vert z(t)\rangle &=& e^{\frac{-i\hat{H} t}{\hbar}}\vert z(0)\rangle\nonumber\\ 
&=& e^{\frac{-i\hat{H} t}{\hbar}}e^{-\frac{|z(0)|^2}{2}}\sum _{n=0}^\infty \frac{z^n(0)}{\sqrt{n!}}|n\rangle \nonumber\\
&=& e^{-\frac{|z(0)|^2}{2}}\sum _{n=0}^\infty \frac{z^n(0)}{\sqrt{n!}}e^{[({-i(n+\frac{1}{2}) \omega t}]}\vert n\rangle\nonumber\\
&=& e^{-\frac{1}{2}\vert z(0)\vert ^2}\sum _{n=0}^\infty\frac{[z(0)e^{-i\omega t}]^n}{\sqrt{n!}}e^{-i\frac{\omega t}{2} }\vert n\rangle\nonumber\\
&=& e^{-i\frac{\omega t}{2} }\vert z(0) e^{-i\omega t}\rangle
\end{eqnarray}
Under time evolution the new state 
$\vert z(t)\rangle = \vert z(0)exp(-i\omega t)\rangle$ 
is also a coherent state apart from a phase 
factor. It means coherent state remains coherent under time evolution. Now we see under the action of annihilation operator on coherent state how the eigen-values change with respect to time.
We know that 
\begin{eqnarray}
\hat{a}\vert z(0)\rangle &=& z(0)\vert z(0)\rangle\nonumber\\
\hat{a}\vert z(t)\rangle &=& z(t)\vert z(t)\rangle\nonumber
\end{eqnarray}
where, $z(t) = z(0)e^{-i\omega t}$.\\
Now we associate $z(t)$  with phase space variable $q(t)$, $p(t)$ as follows:
\begin{equation}
z(t) =  q(t)+ip(t)   
\end{equation}
It is observed that the time propagation of complex function $z(t)$  corresponds to the motion of linear harmonic oscillator.
 \begin{eqnarray}
 z(t) &=& z(0)exp(-i\omega t) \nonumber\\
or, q(t) + ip(t) &=& [q(0) + ip(0)]exp(-i\omega t) \nonumber\\
or, q(t) + ip(t) &=& [q(0) + ip(0)] [cos\omega t - isin\omega t] \nonumber
 \end{eqnarray}
Therefore,\\ 
\begin{subequations}
\begin{equation}
 q(t) = q(0) cos\omega t + p(0) sin\omega t   
\end{equation}
\begin{equation}
p(t) = p(0) cos\omega t - q(0) sin\omega t  
\end{equation}
\end{subequations} 
From the above two equations we find the time evolution of complex function inherently corresponds to simple harmonic motion of linear harmonic oscillator.\\
Now we conclude that the single mode continuous beam of light 
corresponds to a one dimensional linear harmonic oscillator. This connection is purely mathematical but logical. The algebraic connection between linear harmonic oscillator and quantized electromagnetic beam is one to one. It is worthy to mention that the algebra for the two different systems are same but at the interpretation level it is quite different. For example $\vert n\rangle$ represents $n$ photon state in case of single mode continuous beam of light and for linear harmonic oscillator it is interpreted as $n$-th excited state.
In the next section we will discuss coherent states.
\chapter{Coherent State}
\vspace{0.3cm}
We need coherent state representation to interpret the state of a statistical beam of photons (i.e. electromagnetic wave). This state brings up a close relationship between the quantum and classical correlation functions.\\
Coherent state is an eigen state of photon annihilation operator $\hat{a}$. Here $\hat{a}$ is not hermitian operator, and then eigen value is in general complex number. 
\begin{equation}
  \hat{a}\vert z\rangle = z \vert z\rangle \nonumber  
\end{equation}
 $z$ is in general complex number. Then we must have
 \begin{equation}
 \langle z\vert \hat{a}^{\dag} = z^{*} \langle z \vert\nonumber
 \end{equation}
\section{Fock state representation of coherent state}
\vspace{0.3cm}
Coherent state can be can be generated from vacuum state $\vert 0\rangle$ by operating displacement operator on it. Let the displacement operator be
\begin{equation}
 \hat{D}(z)=e^{z\hat{a}^{\dag}-z^{*}\hat{a}}  
\end{equation}
Then the coherent state is
\begin{eqnarray}
\vert z\rangle &=& \hat{D}(z)\vert 0\rangle\nonumber\\ 
or, \vert z\rangle &=& e^{z\hat{a}^{\dag}-z^{*}\hat{a}} \vert 0\rangle \nonumber\\
0r, \vert z\rangle &=& e^{z\hat{a}^{\dag}}e^{-z^{*}\hat{a}}e^{-\frac{\vert z\vert^2}{2}}\vert 0\rangle\nonumber\\
or, \vert z\rangle &=& e^{-\frac{\vert z\vert ^2}{2}}e^{z\hat{a}^{\dag}}\vert 0\rangle \nonumber
\end{eqnarray}
as $e^{-z^{*}\hat{a}}\vert 0\rangle = e^{-z*0}\vert 0\rangle = 1\vert 0\rangle$
\begin{eqnarray}
\vert z\rangle &=& e^{-\frac{\vert z\vert ^2}{2}}\sum_{n=0}^\infty \frac{z^n}{n!}{\hat{a}^{\dag n}}\vert 0\rangle\nonumber\\
or, \vert z\rangle &=& e^{-\frac{\vert z\vert^2}{2}}\sum_{n=0}^\infty \frac{z^n}{\sqrt{n!}}\vert n\rangle 
\end{eqnarray}
as $\vert n\rangle = \frac{\hat{a}^{\dag n}}{\sqrt{n!}}\vert 0\rangle$\\
The above equation represents the number state representation of coherent state. This equation can also be evaluated in another way.We know that the Fock states$\vert n\rangle$ form a complete state and the coherent state $\vert z\rangle$ can be represented as 
\begin{equation}
  \vert z\rangle =\sum_{n=0}^\infty c_n \vert n\rangle 
\end{equation}
where $c_n$ is ingeneral complex. Now we find $c_n$'s.
\begin{eqnarray}
\hat{a}\vert z\rangle &=& z \vert z\rangle \nonumber \\
or, \hat{a}\sum_{n=0}^\infty c_n \vert n\rangle &=& z\sum_{n=0}^\infty c_n \vert n\rangle \nonumber\\
or, \sum_{n=1}^\infty c_n\sqrt{n}\vert n-1\rangle &=& z\sum_{n=0}^\infty c_n \vert n\rangle
\end{eqnarray}
Since, $\vert n\rangle$ are mutually orthogonal set of vectors, this equation is satisfied only if coefficients of Fock space vectors on both sides are equal. Now equating the coefficients of $\vert n-1\rangle$ we get
\begin{eqnarray}
c_n \sqrt{n} &=& zc_{n-1} \nonumber\\
or, c_n &=& \frac{z}{\sqrt{n}}c_{n-1}
\end{eqnarray}
This is a recursion relation which connect the coefficients $c_n$'s. Now by using this relation we obtain 
\begin{eqnarray}
c_{n-1} &=& \frac{z}{\sqrt{n-1}}c_{n-2}  \mbox{ and }\\
c_n &=& \frac{z}{\sqrt{n}}\frac{z}{\sqrt{n-1}}c_{n-2} \nonumber\\
or, c_n &=& \frac{z^2}{\sqrt{n(n-1)}}c_{n-2}.
\end{eqnarray}
Successively substituting we get 
\begin{eqnarray}
c_n &=& \frac{z^2}{\sqrt{n(n-1)(n-2)......2.1}}c_0 \nonumber\\
or, c_n &=& \frac{z^n}{\sqrt{n!}}c_0
\end{eqnarray}
Therefore, the coherent state can be written as
\begin{equation}
  \vert z\rangle = c_0\sum_{n=0}^\infty \frac{z^n}{\sqrt{n!}}\vert n\rangle  
\end{equation}
where, $n = 0,1,2,....$ and 
$c_0$ is a constant,it can be determined by normalizing $\vert z\rangle$. Thus,
\begin{eqnarray}
\langle z\vert z \rangle = 1\nonumber\\
or, \vert c_0 \vert^2\sum_{n=0}^\infty \sum_{m=0}^\infty\frac{z^{*n}{z^m}}{\sqrt{n!m!}}\langle n\vert m\rangle =1\nonumber\\
or, \vert c_0\vert ^2\sum_{n=0}^\infty \sum_{m=0}^\infty\frac{z^{*n}{z^m}}{\sqrt{n!m!}}\delta_{n,m} = 1\nonumber\\
or, \vert c_0\vert ^2\sum_{n=0}^\infty\frac{\vert z\vert^{2n}}{n!} = 1 \nonumber\\
or,  c_0 = e^{-\frac{|z|^2}{2}}e^{i\alpha}
\end{eqnarray}
Therefore,apart from a phase factor the Fock state representation of coherent state can be written as
\begin{equation}
  \vert z\rangle = e^{-\frac{{\vert z\vert}^2}{2}}\sum_{n=0}^\infty \frac{z^n}{\sqrt{n!}}\vert n\rangle  
\end{equation}
where, $n=0,1,2,....$ and
\begin{equation}
 \langle z\vert =e^{-\frac{{\vert z\vert}^2}{2}}\sum_{n=0}^\infty \frac{z^{*n}}{\sqrt{n!}}\langle n\vert   
\end{equation}
where, $n=0,1,2,....$\\
From the above representation it is clear that the projection of $\vert z\rangle$ on every Fock state gives a nonzero value for for all nonzero complex number $z$.
Thus,
\begin{equation}
  \langle n\vert z\rangle = e^{-\frac{{\vert z\vert}^2}{2}}\frac{z^n}{\sqrt{n!}}
\end{equation}
when, $z = 0$, the coherent state $\vert z\rangle$ becomes vacuum state $\vert 0\rangle$ which may be considered as either a coherent state or a Fock state. Here, $P(n) = \vert \langle n\vert z\rangle\vert^2$ represents the probability that $n$ excitations or photons will be found in the coherent state $\vert z\rangle$.
Thus,
\begin{equation}
  P(n) = \vert \langle n\vert z\rangle\vert^2 = e^{-{\vert z\vert^2}}\frac{\vert z\vert^{2n}}{n!}
\end{equation}
The average number of photons for the state $\vert z\rangle$ is found to be 
\begin{eqnarray}
\Bar{n} &=& \sum_{n=0}^\infty nP(n) \nonumber\\
&=&\sum_{n=0}^\infty n e^{-|z|^2} \frac{|z|^{2n}}{n!} \nonumber\\
&=&|z|^2 e^{-|z|^2}\sum_{n=1}^\infty \frac{|z|^{2(n-1)}}{(n-1)!}\nonumber\\
&=& |z|^2 e^{-|z|^2}e^{|z|^2} \nonumber\\
&=& \langle z\vert \hat{a}^{\dag}\hat{a}\vert z\rangle \nonumber
\end{eqnarray}
\begin{equation}
 \therefore \Bar{n}=\sum_{n=0}^\infty nP(n)=\langle z|\hat{a}^{\dag}\hat{a}|z\rangle=|z|^2  
\end{equation}
The average number of photons $\Bar{n}$ depend on $|z|$. So, the $\Bar{n}$ of photon is large or small as the $|z|$ value is either large or small. But, no matter how small $|z|$ may be but $\Bar{n}$  always give a non zero value of probability $P(n)$ except when $|z|=0$.\\
We can write the probability $P(n)$ in terms of $\Bar{n}$ i.e.
\begin{equation}
 P(n)=e^{-\Bar{n}}\frac{\Bar{n}^n}{n!}   
\end{equation}
 which is a Poisson distribution.\\
The remarkable feature about coherent state is that the state remains unchanged even when the annihilation operator acts on it. It is obvious also in some sense as the coherent state is a sum of infinite number of Fock states each of which has different occupation number of photons. So, the entire sum should have occupying a large number of photons similar to a statistical beam of light and so annihilating one or more will not really change the total number of photons. This kind of property of coherent state it  implies a connection between quantum and classical fields. It suggests that it is possible to absorb photons repeatedly from electromagnetic field in coherent state without changing the state in any way. An eigen state of the annihilation operator i.e. coherent state has a Poissonian number distribution when expressed in a basis of energy eigen states as shown above.\\
To verify $\hat{a}\vert z\rangle=z\vert z\rangle$, we substitute $\vert z\rangle$ by the photon number state representation as shown in the above 
\begin{eqnarray}
\hat{a}\vert z\rangle &=& e^{-\frac{\vert z\vert ^2}{2}}\sum_{n=0}^\infty \frac{z^n}{\sqrt{n!}}\hat{a}\vert n\rangle\nonumber\\
or, \hat{a}\vert z\rangle &=& e^{-\frac{\vert z\vert^2}{2}}\sum_{n=0}^\infty \frac{z^n}{\sqrt{n!}}\sqrt{n}\vert{n-1}\rangle\nonumber\\
or, \hat{a}\vert z\rangle&=& z e^{-\frac{\vert z\vert ^2}{2}}\sum_{n=1}^\infty \frac{z^{n-1}}{\sqrt{(n-1)!}}\vert {n-1}\rangle\nonumber\\ 
or, \hat{a}\vert z\rangle &=& z e^{-\frac{|z|^2}{2}}\sum_{n=0}^\infty \frac{z^n}{\sqrt{n!}}|n\rangle \nonumber\\
or, \hat{a}\vert z\rangle &=& z \vert z\rangle \nonumber
\end{eqnarray}
It is verified that coherent state $\vert z\rangle$ is a right eigen state of annihilation operator $\hat{a}$ with an eigen value $z$.\\
Dual to the equation $\hat{a}\vert z\rangle=z \vert z\rangle$ is $\langle z\vert \hat{a}^{\dag}=z^{*}\langle z\vert$. Now we can say that $\langle z\vert$ is the left eigen state of $\hat{a}^{\dag}$ with an eigen value $z^*$.
\section{$\hat{a}^{\dag}$ does not have right eigen state}
\vspace{0.2cm}
It can be proved that it is impossible to find an eigen state $\vert \vert z\rangle\rangle$ of creation operator $\hat{a}^{\dag}$ with a finite eigen value $z$.\\
For the time being we consider 
\begin{equation}
  \hat{a}^{\dag} \vert \vert z\rangle\rangle = z \vert \vert z \rangle\rangle
\end{equation}
and its dual is 
\begin{equation}
 \langle\langle z \vert \vert\hat{a} =\langle\langle z \vert \vert z^{*}   
\end{equation}
Now we write 
\begin{equation}
  \vert \vert z\rangle\rangle = \sum_{n=0}^\infty |n\rangle\langle n\vert \vert z\rangle\rangle  
\end{equation}
 as $\sum_{n=0}^\infty \vert n\rangle\langle n\vert =\hat{I}$ \\
We know $\hat{a}\vert n\rangle = \sqrt{n}\vert n-1\rangle$. If we operate $\hat{a}$ from left successively we obtain
\begin{eqnarray}
\hat{a}^2\vert n\rangle = \sqrt{n(n-1)}|n-2\rangle \nonumber\\
............................................ \nonumber\\
\hat{a}^n\vert n\rangle = \sqrt{n(n-1).....1}|0\rangle\nonumber \\
or, \hat{a}^n\vert n\rangle=\sqrt{n!}\vert 0\rangle
\end{eqnarray}
Its dual is 
\begin{equation}
 \langle n\vert \hat{a}^{\dag n}=\sqrt{n!}\langle 0\vert  
\end{equation}
Now, operating $\vert \vert z\rangle\rangle $ from right in the above equation we find
\begin{eqnarray}
\langle n\vert \hat{a}^{\dag n}\vert \vert z\rangle\rangle &=& \sqrt{n!}\langle 0\vert \vert z\rangle\rangle\nonumber\\
or, z^n\langle n\vert \vert z\rangle\rangle &=& \sqrt{n!}\langle 0\vert \vert z\rangle\rangle\nonumber\\
or, \langle n\vert \vert z\rangle\rangle &=& \frac{\sqrt{n!}}{z^n}\langle 0\vert \vert z\rangle\rangle
\end{eqnarray}
Substituting $\langle n\vert \vert z\rangle\rangle$ in Eq.(76), we find 
\begin{equation}
  \vert \vert z\rangle\rangle = \langle 0\vert \vert z\rangle\rangle \sum_{n=0}^\infty \frac{\sqrt{n!}}{z^n} \vert n\rangle 
\end{equation}
Now the norm square of $\vert \vert z\rangle\rangle$ is found to be 
\begin{equation}
\langle\langle z \vert \vert z\rangle \rangle = \vert \langle   0\vert z\rangle\rangle\vert ^2\sum_{n=0}^\infty \frac{n!}{\vert z\vert^{2n}}
\end{equation}
\hspace{3.5cm} as $\langle n\vert m \rangle =\delta_{n,m}$\\
This squared norm of $\vert \vert z\rangle\rangle$ is divergent for all finite complex amplitude $z$. So, the states $\vert \vert z\rangle\rangle$ which was previously considered as right eigen states of $\hat{a}^{\dag}$ can't be considered as physically admissible states of radiation field. We need the states $\vert n\rangle, \vert z\rangle$ as the basis to represent the electromagnetic radiation field.
\section{Coherent state representation of Fock state}
\vspace{0.2cm}
Can we express Fock state $\vert n\rangle$ in terms of coherent state $\vert z\rangle$?\\
We know, 
\begin{equation}
   \vert z\rangle = e^{-\frac{|z|^2}{2}}\sum_{n=0}^\infty \frac{z^n}{\sqrt{n!}}\vert n\rangle
\end{equation}
Now multiply the above equation (Eq.(3.25)) by   \\ $\frac{1}{\pi}\frac{1}{\sqrt{m!}}z^{*m}e^{-\frac{|z|^2}{2}}$; then we find 
\begin{equation}
 \frac{1}{\pi}\frac{1}{\sqrt{m!}}z^{*m}e^{-\frac{|z|^2}{2}}|z\rangle=\frac{1}{\pi}e^{-|z|^2}\sum_{n=0}^\infty \frac{{z^n}{z^{*n}}}{\sqrt{n!m!}}|n\rangle   
\end{equation}
Integrating both sides over z 
\begin{eqnarray}
 \frac{1}{\pi}\int{\frac{1}{\sqrt{m!}}z^{*m}e^{-\frac{|z|^2}{2}}|z\rangle d^2 z}  = \frac{1}{\pi}\sum_{n=0}^\infty \int{ e^{-|z|^2} \frac{{z^n}{z^{*n}}}{\sqrt{n!m!}}|n\rangle d^2} z \nonumber\\
\end{eqnarray}
We use polar coordinates to evaluate the integral of the R.H.S. of the above equation as: $z=re^{i\theta}$ and $d^2 z = r dr d\theta$.
\begin{eqnarray}
  \frac{1}{\pi}\sum_{n=0}^\infty \int_{r=0}^\infty rdr \int_{\theta=0}^{2\pi} d\theta 
e^{-r^{2}} \frac{r^{n+m}e^{i(n-m)\theta}}{\sqrt {n!m!}}|n\rangle \nonumber\\
= \frac{1}{\pi} \times 2\pi \times \frac{1}{n!} \int_{r=0}^\infty e^{-r^2}r^{2n+1}dr |n\rangle \nonumber
\end{eqnarray}
\hspace{1.5cm}since, $\int_{\theta=0} ^{2\pi} e^{i(n-m)\theta}d\theta = 2\pi\delta_{n,m}$\\
Again, using the properties of the gamma function,\\
$\int_{y=0}^\infty e^{-y^m}y^n dy=\frac{1}{m}\Gamma(\frac{n+1}{m})$, in the above equation we obtain\\
\begin{eqnarray}
\frac{1}{\pi}\int{\frac{1}{\sqrt{n!}}z^{*n}e^{-\frac{\vert z\vert^2}{2}}\vert z\rangle d^2 z}&=&2. \frac{1}{n!}.\frac{1}{2}\Gamma\left(\frac{2n+2}{2}\right)\vert n\rangle \nonumber\\
or, \frac{1}{\pi}\int{\frac{1}{\sqrt{n!}}z^{*n}e^{-\frac{|z|^2}{2}}|z\rangle d^2 z}&=&\frac{1}{n!}\Gamma(n+1)|n\rangle \nonumber\\
or, \frac{1}{\pi}\int{\frac{1}{\sqrt{n!}}z^{*n}e^{-\frac{|z|^2}{2}}|z\rangle d^2 z}&=& |n\rangle \nonumber
\end{eqnarray}
\hspace{3cm} since, $\Gamma(n+1)=n!$
\begin{equation}
   \therefore |n\rangle=\frac{1}{\pi}\int{\frac{1}{\sqrt{n!}}z^{*n}e^{-\frac{|z|^2}{2}}|z\rangle d^2 z} 
\end{equation}
\section{Properties of Coherent state}
\subsection{\textit{Property-I:} Possesses Minimum Uncertainty product}
Consider two non commuting observables $\hat{A}$ and $\hat{B}$, and the state of the system is described with suitably normalized state ket, $\vert \psi\rangle$, then the uncertainty product of these operators can be written as $\Delta \hat{A} \Delta \hat{B}\geqslant \frac{1}{2}\vert\langle[\hat{A}, \hat{B}]\rangle \vert$. Here we wish to calculate the position momentum uncertainty product with respect to coherent state and show that it will satisfy minimum uncertainty relation. Using the definition of coherent state, we can write $\hat{a}\vert z\rangle = z\vert z\rangle$ and $\hat{a}^{\dag}\vert z\rangle = z^{*}\vert z\rangle$ and we have the position and momentum operator in terms of creation and annihilation operators i.e. $\hat{x}=\sqrt{\frac{\hbar}{2m\omega}}(\hat{a}^{\dag}+\hat{a})$ and $\hat{p}_x=i \sqrt{\frac{m\hbar\omega}{2}}(\hat{a}^{\dag}-\hat{a})$. The uncertainty in measurement of position with respect to coherent state is 
\begin{eqnarray}
(\Delta x)^2_{\vert z\rangle} &=& \langle \hat{x}^2\rangle_{\vert z\rangle} - \langle \hat{x}\rangle^2_{\vert z\rangle} \nonumber\\
(\Delta x)^2_{\vert z\rangle} &=& \frac{\hbar}{2m\omega}[\langle z\vert (\hat{a}^{\dag}+\hat{a})^2 \vert z\rangle - {\langle z\vert (\hat{a}^{\dag}+\hat{a}) \vert z\rangle}^2] \nonumber\\
\end{eqnarray}
Now,
$\langle z|(\hat{a}^{\dag}+\hat{a})|z\rangle =z^*+z $ 
and;
\begin{eqnarray}
\langle z|(\hat{a}^{\dag}+\hat{a})^2|z\rangle &=&\langle z|\hat{a}^{\dag}\hat{a}^{\dag}+\hat{a}^{\dag}\hat{a}+\hat{a}\hat{a}^{\dag}+\hat{a}\hat{a}|z\rangle\nonumber\\
&=& \langle z|\hat{a}^{\dag}\hat{a}^{\dag}+2\hat{a}^{\dag}\hat{a}+\hat{I}+\hat{a}\hat{a}|z\rangle \nonumber\\
&=& z^{*2}+2zz^*+1+z^2 \nonumber\\
&=& 1+(z+z^*)^2
\end{eqnarray}
Finally,
\begin{eqnarray}
(\Delta x)^2_{\vert z\rangle} &=& \frac{\hbar}{2m\omega}[1+(z+z^*)^2-(z+z*)^2] \nonumber\\
&=& \frac{\hbar}{2m\omega} \nonumber
\end{eqnarray}
\begin{equation}
   \therefore (\Delta x)_{|z\rangle}=\sqrt{\frac{\hbar}{2m\omega}}
\end{equation}
Now, let us calculate the uncertainty in measurement of momentum with respect to coherent state i.e.
\begin{eqnarray}
(\Delta p_x)^2_{\vert z\rangle} &=& {\langle \hat{p}_x^2\rangle}_{\vert z\rangle}-{{\langle \hat{p}_x\rangle}}^2_{\vert z\rangle} \nonumber\\
or, (\Delta p_x)^2_{\vert z\rangle} &=& -\frac{\hbar m \omega}{2}[\langle z\vert (\hat{a}^{\dag}-\hat{a})^2| z\rangle-{\langle z\vert (\hat{a}^{\dag}-\hat{a})\vert z\rangle}^2]\nonumber\\
\end{eqnarray}
Now, $\langle z\vert (\hat{a}^{\dag}-\hat{a})\vert z\rangle = z^*-z $ and 
\begin{eqnarray}
\langle z\vert (\hat{a}^{\dag}-\hat{a})^2\vert z\rangle &=& \langle z\vert \hat{a}^{\dag}\hat{a}^{\dag}-\hat{a}^{\dag}\hat{a}-(\hat{I}+\hat{a}^{\dag}\hat{a})+\hat{a}\hat{a}|z\rangle\nonumber\\
&=& \langle z\vert \hat{a}^{\dag}\hat{a}^{\dag}-2\hat{a}^{\dag}\hat{a}-\hat{I}+\hat{a}\hat{a}\vert z\rangle\nonumber\\
&=& z^{*2}-2zz^*-1+z^2 \nonumber\\
&=& (z^*-z)^2-1
\end{eqnarray}
So, 
\begin{eqnarray}
(\Delta p_x)^2_{|z\rangle}&=&-\frac{\hbar m \omega}{2}[(z^*-z)^2-1-(z^*-z)^2]
\nonumber\\
&=&\frac{\hbar m \omega}{2} \nonumber
\end{eqnarray}
\begin{equation}
    (\Delta p_x)_{|z\rangle}=\sqrt{\frac{\hbar m\omega}{2}}
\end{equation}
\begin{equation}
    \therefore (\Delta x)_{|z\rangle}(\Delta p_x)_{|z\rangle}=\sqrt{\frac{\hbar}{2m\omega}}\sqrt{\frac{\hbar m\omega}{2}}=\frac{\hbar}{2}
\end{equation}
It is shown that coherent state satisfies Heisenberg's minimum uncertainty product.
\subsection{\textit{Property-II:} Completeness relation}
Though coherent states does not follow the orthonormality condition  but the other crucial property of coherent states as a basis set is the completeness relation. It is important to note that orthonormality is a scalar condition, while completeness is an operator relation.\\
Here the idea is that there is a continuous basis {$\vert z\rangle$} in the linear vector space, labelled by the values of ${z}$. The elements of this basis satisfy completeness relation 
\begin{equation}
 \frac{1}{\pi}\int d^2 z \vert z\rangle\langle z\vert = \hat{I}    
\end{equation}
Starting from left hand side we try to get right hand side of the relation. Now we represent $\vert z\rangle$, $\langle z\vert$ in Fock basis.
\begin{equation}
  \int d^2 z \vert z\rangle\langle z\vert =\int d^2 
z e^{-\vert z\vert^2}\sum_{n=0}^\infty\sum_{m=0}^\infty \frac{z^n z^{*m}}{\sqrt{n!m!}}\vert n\rangle \langle m\vert  
\end{equation}
Using plane polar coordinates: $z = re^{i\theta}$ and $d^2 z = r dr d\theta$ we obtain 
\begin{eqnarray}
\int d^2 z \vert z\rangle\langle z\vert
 = \hspace{6cm}\nonumber\\ 
 \sum_{n=0}^\infty\sum_{m=0}^\infty \frac{\vert n\rangle \langle m\vert}{\sqrt{n!m!}}\int_{\theta=0}^{2\pi}d\theta e^{i(n-m)\theta}\int_{r=0}^\infty dr re^{-r^2}r^{n+m}\nonumber\\
\end{eqnarray}
Since, $\int_{\theta = 0}^{2\pi}d\theta e^{i(n-m)\theta} = 2\pi\delta_{n,m}$; then 
\begin{equation}
   \int d^2 z \vert z\rangle\langle z\vert =
  \sum_{n=0}^\infty\frac{\vert n\rangle \langle n\vert}{n!}2\pi\int_{r=0}^\infty dr e^{-r^2}r^{2n+1}
\end{equation}
Using the property of Gamma function: 
$\int_{y=0}^\infty dy e^{-y^m}y^n=\frac{1}{m}\Gamma(\frac{n+1}{m})$
we obtain
\begin{eqnarray}
\int d^2 z \vert z\rangle\langle z\vert &=&\sum_{n=0}^\infty\frac{|n\rangle \langle n|}{n!}2\pi\frac{1}{2}\Gamma(\frac{2n+2}{2})\nonumber\\
 &=& \sum_{n=0}^\infty\frac{|n\rangle \langle n|}{n!}\pi n!; as \Gamma(n+1)=n! \nonumber\\
 &=&\pi\sum_{n=0}^\infty
\vert n\rangle \langle n\vert \nonumber
\end{eqnarray}
\begin{eqnarray}
\frac{1}{\pi}\int d^2 z \vert z\rangle\langle z\vert &=& \sum_{n=0}^\infty\vert n\rangle \langle n\vert  \nonumber\\
or, \frac{1}{\pi}\int d^2 z \vert z\rangle\langle z\vert &=& \hat{I}
\end{eqnarray}
Now we see that the elements of coherent state basis satisfy the completeness relation.
\subsection{\textit{Property-III:} Non orthogonality and Over completeness }
Now,let us verify whether the coherent states are orthogonal or not. Consider ant two coherent states, say $\vert z_1\rangle$ and $\vert z_2\rangle$. The Fock state representation of these two states is 
\begin{eqnarray}
\vert z_1\rangle &=& e^{-\frac{\vert z_1\vert^2}{2}} \sum_{n=0}^{\infty} \frac{z_1^n}{\sqrt{n!}}\vert n\rangle \\
\vert z_2\rangle &=& e^{-\frac{\vert z_2\vert^2}{2}}\sum_{m=0}^{\infty} \frac{z_2^m}{\sqrt{m!}} \vert m\rangle
\end{eqnarray}
Now taking the scalar product of these two state kets, we get
\begin{eqnarray}
\langle z_1\vert z_2\rangle &=& \left (e^{-\frac{\vert z_1\vert ^2}{2}}\sum_{n=0}^{\infty} \frac{z_1^{*n}}{\sqrt{n!}}\langle n\vert \right)\left( e^{-\frac{\vert z_2\vert^2}{2}}\sum_{m=0}^{\infty} \frac{z_2^m}{\sqrt{m!}}\vert m\rangle\right) \nonumber\\
&=& e^{-\frac{\vert z_1\vert^2}{2}}e^{-\frac{\vert z_2\vert ^2}{2}}\sum_{n=0}^\infty\sum_{m=0}^{\infty} \frac{z_1^{*n}}{\sqrt{n!}}{\frac{{z_2}^m}{\sqrt{m!}}}\langle n|m\rangle \nonumber\\
&=& e^{-\frac{|z_1|^2}{2}}e^{-\frac{|z_2|^2}{2}}\sum_{n=0}^\infty{\frac{{{z_1}^*}^n{z_2}^n} {n!}}\nonumber\\
&=& e^{-\frac{|z_1|^2}{2}}e^{-\frac{|z_2|^2}{2}}e^{{z_1}^*z_2}
\end{eqnarray}
Similarly,
\begin{equation}
  \langle z_2\vert z_1\rangle = e^{-\frac{\vert z_1|^2}{2}}e^{-\frac{|z_2|^2}{2}}e^{{z_1}
{z_2}^*}  
\end{equation}
We see that these two states are non orthogonal to each other.Hence coherent states are non orthogonal.Now, we can calculate the projection probability of $\vert z_1\rangle$ on $\vert z_2\rangle$ i.e.
\begin{eqnarray}
\vert \langle z_1 \vert z_2\rangle\vert^2 &=& \langle z_1\vert z_2\rangle \langle z_2\vert z_1\rangle\nonumber\\
&=& e^{-\vert z_1\vert^2 -\vert z_2\vert^2 +{z_1}^* z_2 +z_1{z_2}^*}\nonumber\\
&=& e^{-\vert z_1 -z_2\vert^2}
\end{eqnarray}
It is seen that lower projection probability can be achieved when the states are further apart in phase space. Note that $\vert z_1\rangle$ and $\vert z_2\rangle$ are approximately orthogonal when $\vert z_1 -z_2\vert$ becomes large. One can obtain zero projection probability when $\vert z_1\rangle$, $\vert z_2\rangle$ are apart by infinity.\\
This also indicates that if a system is in the state $\vert z_1\rangle$, then there is a non zero probability that the system will be found in any other coherent state $\vert z_2\rangle$. In other words it is also said that coherent states are linearly dependent. It means one coherent state can be represented in terms of other states. So, let us do this
\begin{eqnarray}
    \vert z^{'}\rangle &=& \hat{I} \vert z^{'} \rangle \nonumber\\
    &=& \frac{1}{\pi} \int \vert z\rangle\langle z \vert z^{'} \rangle d^{2} z \nonumber\\
&=& \frac{1}{\pi}\int \vert z\rangle e^{-\frac{\vert z-z^{'}\vert ^{2}}{2}} e^{\frac{z^{*}z^{'}-zz^{'*}}{2}}d^2z
\end{eqnarray}
Therefore one can say that any one coherent state can be represented in terms of all of them. Above all we can say that though coherent states are all normalized but not orthogonal; but they are complete in the sense that they furnish a resolution of identity. Actually they form an over complete basis set.\\\\
\textbf{What does over completeness mean?}\\\\
In usual sense a basis set or basis in a Linear vector space (LVS) is a set of vectors in the LVS satisfying two requirements: (i) Linear independence (ii) Spanning the space. The elements of this basis satisfy the orthonormality and completeness relations. Now one can suggest the idea to consider set of coherent states as a continuous basis $\vert z\rangle$ in the LVS although no two coherent states are orthogonal but they satisfy completeness relation. Any coherent state can be described by other set of coherent states. This is in sharp contraction with the requirement of linear independence. Most surprisingly it is seen that a subset of coherent state basis vectors can furnish the resolution of identity operator. This is impossible in the context of position or momentum basis.\\
Properties of over completeness can easily be illustrated qualitatively following the book by Mandel and Wolf \cite{Mandel}. Let's consider a vector $\vec{r}$ as the position vector of a point in $R^2$ vector space. $\hat{e}_1$, $\hat{e}_2$ are the two orthonormal vectors. We can represent $\vec{r}$ as 
\begin{equation}
    \vec{r} = c_1 \hat{e}_1+ c_2 \hat{e}_2 
\end{equation}
This set of components $c_1$, $c_2$ are unique.\\
If we consider two non orthogonal unit vectors $\hat{e}_1^{'}$, $\hat{e}_2^{'}$ as basis then the same position vector $\vec{r}$ can be represented as 
\begin{equation}
    \vec{r} = c_{1}^{'} \hat{e}_1^{'}+ c_{2}^{'} \hat{e}_2^{'}
\end{equation}
and the components are also unique as earlier. \\
But if we consider three non orthogonal unit vectors $\hat{e}_1^{''}, \hat{e}_2^{''}, \hat{e}_3^{''}$ as  basis although dependent then we can represent 
\begin{equation}
   \vec{r} = c_{1}^{''} \hat{e}_1^{''} + c_{2}^{''} \hat{e}_2^{''} + c_ {3}^{''} \hat{e}_3^{"} 
\end{equation}
The set of components $c_{1}^{''}$, $c_{2}^{''}$, $c_{3}^{''}$ are not unique. Uniqueness is lost here.\\
Now we conclude that if in a LVS; there are more number of basis vectors than needed and they are dependent and also mutually non orthogonal then we can't represent a vector in a unique way. That basis set is not only complete but it is over complete.\\
Here, in the above example the LVS is finite dimensional and the over completeness property can be removed by the removal of any one of basis vector. But it is not so easy when we deal with the continuous basis like coherent state basis. Because of non orthogonal property of coherent state we have to face some problem to interpret the projection probability of coherent state $\vert z\rangle$ on some other  state $\vert \phi\rangle$. First we try with a Fock state $\vert n\rangle$. Here the scalar product $\langle n\vert \phi\rangle$ represents the probability amplitude of $\vert \phi\rangle$ in Fock basis. $\vert \langle n\vert \phi \rangle\vert^2$ is the probability of finding $n$ photons in state $\vert \phi\rangle$. As the Fock basis vectors are orthogonal to each other and the probabilities $\vert \langle n\vert \phi \rangle\vert^2$ are mutually exclusive for different $n$ so 
\begin{eqnarray}
\sum_{n=0}^\infty\vert \langle n\vert\phi \rangle\vert^2 &=& \sum_{n=0}^\infty \langle \phi\vert n\rangle \langle n\vert \phi\rangle \nonumber\\
&=& \langle \phi\vert \sum_{n=0}^\infty [\vert n\rangle \langle n\vert]\phi\rangle \nonumber\\
&=& \langle\phi\vert \phi\rangle \nonumber\\
&=& 1
\end{eqnarray}
On the other hand, for coherent state $\vert z\rangle$ the square of the scalar product $\vert \langle z\vert \phi \rangle\vert^2 $ do not represent mutually exclusive probabilities and do not integrate to unity. Here it is
\begin{eqnarray}
\int \vert \langle z\vert \phi\rangle\vert^2 d^2 z &=& \int\langle\phi\vert z\rangle \langle z\vert \phi\rangle d^2 z \nonumber\\
&=&\pi
\end{eqnarray}
Mathematically over completeness of basis set implies that even a subset of coherent states may suffice to form a complete set. Following C. L. Mehta's article \cite{Mehta} on \emph{Sudarshan: Seven Science Quests} it can be shown that \emph{even very restricted sub-set of the coherent states, namely, those on the unit circle $(z = e^{i\theta})$ alone form a complete set.}\\
In a complex plane each point denotes a complex number. Point at infinity corresponds to the complex number $(z = re^{i\theta})$ with $r\in [0,\infty)$, for all values of $\theta$. The finite part of the complex plane together with the point at infinity, i.e.
$C U\{\infty\}$, is called the extended complex plane. It is denoted by $\hat{C}$.\\
For our purpose we take a very restricted sub-set of complex numbers, those on the unit circle. Now, we associate each complex number $e^{i\theta}$, $0 < \theta \leqslant 2\pi$ with base ket $\vert e^{i\theta}\rangle$. Obviously,
$\vert e^{i\theta}\rangle$ gives a subset of coherent states. Now it is shown that this subset can furnish the identity operator $\hat{I}$. Calculate 
$\int d\theta
\vert e^{i\theta}\rangle\langle e^{i\theta}\vert$ using the Fock basis representation of $\vert e^{i\theta}\rangle\langle e^{i\theta}\vert$.
\begin{eqnarray}
\int d\theta\vert e^{i\theta}\rangle\langle e^{i\theta}\vert &=& e^{-1}{\sum_{n=0}^\infty}{\sum_{m=0}^\infty}
\int d\theta {\frac{e^{i(n-m)\theta}}{\sqrt{n!m!}}}\vert n\rangle \langle m \vert \nonumber\\
&=& \frac{2\pi}{e}\sum_{n=0}^\infty \frac{\vert n\rangle \langle n\vert}{n!} 
\end{eqnarray}
\begin{eqnarray}
\therefore \frac{e}{2\pi}(\hat{a}^{\dag}\hat{a})!\int d\theta\vert e^{i\theta}\rangle\langle e^{i\theta}\vert &=&\sum_{n=0}^\infty |n\rangle \langle n| \nonumber\\
or, \frac{e}{2\pi}(\hat{a}^{\dag}\hat{a})!\int d\theta\vert e^{i\theta}\rangle\langle e^{i\theta}\vert &=&\hat{I}
\end{eqnarray}
It suggests that the subset of coherent states can also furnish the resolution of identity operator in a funny way.\\
We illustrate the over-completeness property by observing that the $\vert n\rangle$ may be written using the subset of coherent states $\{\vert e^{i\theta}\rangle\}$.
\begin{equation}
   \vert e^{i\theta}\rangle = e^{-\frac{1}{2}}\sum_{n=0}^\infty\frac{e^{in\theta}}{\sqrt{n!}}\vert n\rangle 
\end{equation}
Multiply both sides by $\frac{1}{\pi\sqrt{m!}}e^{-im\theta}e^{-\frac{1}{2}}$
and integrating we obtain
\begin{eqnarray}
\frac{1}{\pi\sqrt{m!}}e^{-\frac{1}{2}}\int e^{-im\theta}\vert e^{i\theta} \rangle d\theta 
&=& \frac{e^{-1}}{\pi}\sum_{n=0}^\infty \int\frac{e^{i(n-m)\theta}}
{\sqrt{n!m!}}\vert n\rangle d\theta \nonumber\\
&=& \frac{e^{-1}2\pi}{\pi m!} \vert m\rangle 
\end{eqnarray}
\begin{eqnarray}
\therefore \frac{1}{\pi e} \frac{2\pi}{m!}\vert m\rangle &=& \frac{e^{-\frac{1}{2}}}{\pi\sqrt{m!}}\int e^{-im\theta}\vert e^{i\theta}\rangle d\theta \\
\vert m\rangle &=& \frac{\sqrt{m!e}}{2\pi}\int e^{-im\theta} \vert e^{i\theta}\rangle d\theta 
\end{eqnarray}
Hence, Fock state $\vert m\rangle$ is written in terms of the subset of coherent state basis $\{ \vert e^{i\theta}\rangle\}$. \\\\
\vspace{0.3cm}
\textbf{More on over complete basis set:}\\
Consider the matrix elements of an operator $\hat{A}$ in Fock basis. The matrix elements are $\langle n\vert \hat{A}\vert m\rangle$; where, $n, m = 0, 1, 2, 3,...\infty$ exhaust all diagonal and non diagonal matrix elements. It is seen that all the diagonal and non diagonal matrix elements expressed in Fock basis can be written down in coherent state basis using only diagonal term i.e. $\langle z \vert \hat{A}\vert z\rangle$ as 
\begin{equation}
    \langle n\vert \hat{A}\vert m\rangle = \frac{1}{\sqrt{n!m!}}\left[\frac{\partial^{n+m}}{\partial z^{*n} \partial z^m} \langle z\vert \hat{A}\vert z\rangle e^{\vert z\vert^2}\right]_{z^{*},z=0}
\end{equation}
This suggests that only diagonal matrix elements in coherent state basis are enough to write all the diagonal and non diagonal matrix elements in Fock basis. This is the essence of over completeness property of coherent state basis.\\
Let's take an example for $n=2, m=3$. So, the corresponding matrix element is $\langle 2\vert \hat{A}\vert 3\rangle$. Now from the expression  above we have 
\begin{eqnarray}
\frac{1}{\sqrt{n!m!}}\left[\frac{\partial^{n+m}}{\partial z^{*n} \partial z^m}\langle z \vert \hat{A} \vert z\rangle e^{\vert z\vert^2}\right]_{z^{*},z=0} \nonumber
\end{eqnarray}
\begin{eqnarray}
&=& \frac{1}{\sqrt{n!m!}}\left[\frac{\partial^{n+m}}{\partial z^{*n} \partial z^m}e^{-\vert z\vert^2} \sum_{p=0}^{\infty} \sum_{q=0}^{\infty} \frac{z^{*p}z^q}{\sqrt{p!q!}}\langle p\vert \hat{A}\vert q\rangle e^{\vert z\vert ^2}\right]_{z^{*},z=0} \nonumber\\
&=& \frac{1}{\sqrt{n!m!}}\left[\frac{\partial^{n+m}}{\partial z^{*n} \partial z^m} \sum_{p=0}^{\infty} \sum_{q=0}^{\infty} \frac{z^{*p}z^q}{\sqrt{p!q!}} \langle p\vert \hat{A}\vert q\rangle\right]_{z^{*},z=0} \nonumber\\
\end{eqnarray}
Now we put $n=2$ and $m=3$ in the above equation
\begin{equation}
\langle 2\vert \hat{A}\vert 3\rangle =  \frac{1}{\sqrt{2!3!}} \left[\frac{\partial^{2+3}}{\partial z^{*2} \partial z^3} \sum_{p=0}^{\infty} \sum_{q=0}^{\infty} \frac{z^{*p}z^q}{\sqrt{p!q!}} \langle p\vert \hat{A}\vert q\rangle\right]_{z^{*},z=0}
\end{equation}
All the terms come from the summation do not contribute to evaluate the matrix element. Only the term with $p=2$ and $q=3$ survives here. This is so because all the terms with $p<2$ and $q<3$ reduces to zero as the order of differentiation in each is higher than they appear. On the other hand the terms with $p>2$ and $q>3$ vanishes because of the restriction $z^*,z=0$. So the sum in the above expression reduces to 
\begin{eqnarray}
\frac{1}{\sqrt{2!3!}} \left[\frac{\partial^{2+3}}{\partial z^{*2} \partial z^3} \frac{z^{*2}z^3}{\sqrt{2!3!}} \langle 2\vert \hat{A}\vert 3\rangle\right]_{z^{*},z=0} \nonumber
\end{eqnarray}
\begin{eqnarray}
&=& \frac{1}{2!3!}\left[\frac{\partial^{2+3}}{\partial z^{*2} \partial z^3} z^{*2}z^{3}\right]_{z^{*},z=0} \langle 2\vert \hat{A}\vert 3\rangle \nonumber\\
&=& \frac{1}{2!3!} {2!3!} \langle 2\vert \hat{A}\vert 3\rangle \nonumber\\
&=& \langle 2\vert \hat{A}\vert 3\rangle \nonumber
\end{eqnarray}
Now it is verified that any diagonal or non diagonal matrix element of an operator represented in Fock basis can be generated  only using diagonal coherent state basis. This is due to the over 
completeness property of coherent state basis.
\chapter{Readings of E.C.G.  Sudarshan's paper}
\vspace{0.3cm}
Here, we present an elaborate calculation of E. C. G. Sudarshan's famous paper\cite{E.C.G} on quantum optics. It was published in 1963 and titled as \emph{Equivalence of semiclassical equivalence of semiclassical and quantum mechanical descriptions of statistical light beams.}\\
According to this paper \emph{classical theory of optical  coherence.....is adequate for the description of the classical optical phenomena of interference and diffraction in general. More sophisticated experiments on intensity interferometry and photoelectric counting statistics necessitated special higher order correlations. Most of this work was done using a classical or a 
semiclassical formulation of the problem. On the other hand, statistical states of a quantized (electromagnetic) field have been considered recently, and a quantum mechanical definition of coherence functions of arbitrary order presented.}\\
The aim of this paper is to elaborate quantum definition of coherence function and find complete equivalence to classical description.
\section{Classical and Quantum Mechanical Correlation Function}
\vspace{0.3cm}
For the sake of reader, basic definitions and interpretations of correlation functions in classical statistical optics are given.
Classical theory of optical coherence was developed by E. Wolf\cite{Born} by considering optical field as a random processes or as part of a stochastic process. The concept of partial coherence and its propagation laws were clarified and partial coherence was described by the ``two point correlation function", which was physically like the intensity but was propagated like an amplitude.\\
For simplicity we consider only the scalar nature of the fluctuating electric field. Real classical electric field is expressed as
\begin{equation}
    E(x,t) = E^{+}(x,t)+E^{-}(x,t)
\end{equation}
The arguments $x,t$ designates spatial and time coordinates.
$E^{+}(x,t)$ is the complex positive frequency part and 
$E^{-}(x,t)$ $= E^{+*}(x,t)$ complex negative frequency part of real classical electric field respectively.\\
Classical two-point correlation function is defined as the statistical average of the product $E^{-}(x_2,t_2)E^{+}(x_1,t_1)$ of two complex field amplitudes i.e. 
\begin{equation}
    \Gamma^{(1)} = \left<E^{-}(x_2,t_2)E^{+}(x_1,t_1)\right>
\end{equation}
 The averages $\langle(....)\rangle$ are the stochastic averages over the given ensemble. This is adequate to discuss intensity measurements at $x_1 = x_2, t_1 = t_2$. This two-point classical correlation function is measured in Young type interference phenomena.\\
Classical four-point correlation function is defined as 
\begin{equation}
    \Gamma^{(2)} = \left<E^{-}(x_4,t_4)E^{-}(x_3,t_3)E^{+}(x_2,t_2)E^{+}(x_1,t_1)\right> 
\end{equation}
This is needed to discuss Hanbury Brown-Twiss(HBT)\cite{} intensity correlations in a fluctuating classical beam when $(x_4 = x_2, t_4 = t_2), (x_3 = x_1,t_3 = t_1)$.\\
Near about the same time L. Mandel\cite{Mandel} studied photo-electron counting statistics and derived the counting formula for the case when a plane wave of quasi-monochromatic polarized light is incident on a photoelectric detector. The probability $P(n,T)$ that $n$ photo-electrons will be released, in a fixed time interval $T$, is related to the probability density $P(W)$ for the time integrated intensity $W$ by the well known formula as a Poisson transform of $P(W)$, i.e. a linear transform with a Poisson kernel.
\begin{equation}
   P(n,T) = \int_0^\infty P(W)\frac{W^n}{n!}e^{-W}dW 
\end{equation}
Here, 
$W = \alpha \int_{0}^{T} I(t)dt$; $\alpha$ is a measure of the quantum efficiency of the dectector and I(t) is the intensity of the light(measured in photons per second) at time $t$. This work of Mandel helped to understand the HBT effect.\\
In Mandel's treatment of photo electron counting based on the fact that light detection involve absorption of photons from the field being observed. Most features of photo-counting distribution were well analyzed by Mandel's formula. Several effects such as propagation of coherence function, Hanbury Brown Twiss experiments\cite{HBT} on bunching of photons i.e. the photons of a light beam of narrow spectral width have a tendency to arrive in correlated pairs had been adequately explained by classical approach to light fluctuations.\\
With the advent of lasers in 1960, a need arose for the quantum mechanical description of electromagnetic fields associated with arbitrary light beams and its associated coherence functions of arbitrary order.\\
Depending on the methods of light detection due to absorption of photons from the field, Glauber [6] was able to arrive at the most useful measure of (partial) coherence of the quantised electromagnetic field at the two-point level and the generalized it to correlation functions of arbitrary order. In this context it is worthy to mention E.C.G.'s remark again. We highlight here few lines from  the first para of E.C.G.'s paper, \emph{........ More sophisticated experiments on intensity interferometry and photoelectric counting statistics necessitated special higher order correlations. Most of this work was done using a classical or a semiclassical formulation of the problem. On the other hand, statistical states of a quantized (electromagnetic) field have been considered recently, and a quantum mechanical definition of coherence functions of arbitrary order presented.}\\
In phrasing the above para E. C. G. cites R. J. Glauber's paper titled,\emph{Photon Correlations} [6]. In this context it is worthy to mention that the completion of Quantum Electrodynamics and the growth of classical statistical optics merged in the beginning of 1960's that led to the development of quantum theory of optical coherence; partial or complete coherence, to some finite order or to all orders and R. J. Glauber made a significant contribution by introducing coherent state in the center stage of Quantum Optics as a special set of quantum states. These states can be defined both for material oscillators and for the free radiation field.This has been explicitly shown in Sec.2.4.\\
Now, what does it actually mean to correlation functions of quantised electric field? It is the counterpart of classical coherence function.\\
Quantized electric field operator is expressed as
\begin{equation}
  \hat{E}(x,t) = \hat{E}^{+}(x,t) + \hat{E}^{-}(x,t)
\end{equation}
The arguments $x$, $t$ designates spatial and time coordinates.
$\hat{E}^{+}(x,t)$ is the positive frequency (annihilation) part of the field operator and $\hat{E}^{-}(x,t) = \hat{E}^{+\dag}(x,t)$  negative frequency (creation) part of field operator $\hat{E}$ respectively. $\hat{E}^{+}(x,t)$ is an operator which acting on a state annihilates or subtracts one photon; $\hat{E}^{-}(x,t)$ is the hermitian conjugate of $\hat{E}^{+}(x,t)$ acting on a state it creates or adds single photon. In the vacuum state $\vert 0\rangle$ there is no photon at all, so
\begin{equation}
    \hat{E}^{+}(x,t) \vert 0\rangle = \vec{0}
\end{equation}
In brief we say that 
\begin{eqnarray}
\hat{E}^{+} = f(\hat{a}) \nonumber \\
\hat{E}^{-} = f(\hat{a}^{\dag})
\end{eqnarray}
Two-point quantum correlation function is defined to describe intensity measurements by photon absorption,adequate to discuss Young type interference phenomena
\begin{equation}
    G^{(1)} = Tr\left[\hat{\rho}\hat{E}^{-}(x_2,t_2)\hat{E}^{+}(x_1,t_1)\right] 
\end{equation}
where, $\hat{\rho}$ is the density operator of quantum state.\\
Four-point quantum correlation function needed to describe Hanbury- Brown-Twiss intensity correlations
\begin{equation}
    G^{(2)}=Tr\left[\hat{\rho}\hat{E}^{-}(x_4,t_4)\hat{E}^{-}(x_3,t_3)\hat{E}^{+}(x_2,t_2)\hat{E}^{+}(x_1,t_1)\right]
\end{equation}
In the above it is shown that an optical field which is described by a density operator $\hat{\rho}$; then one can define quantum correlation functions of arbitrary order.
\section{Density operator and its representation using diagonal coherent state basis}
\vspace{0.3cm}
States in quantum mechanics can be described by a single state vector or wave function $\psi$ then the state is pure. Here, density matrix $\hat{\rho} = \vert \psi\rangle \langle \psi\vert$; and $\hat{\rho}^2 =\hat{\rho}$ for pure state.\\
But when the state of the system is described by an ensemble of several pure states $\vert \psi_1\rangle, \vert \psi_2\rangle, .. \vert \psi_n\rangle$ each with corresponding probability $p_1, p_1, ...p_n$ respectively then the entire ensemble can be represented by density operator or density matrix
\begin{equation}
    \hat{\rho} = \sum_{i=1}^{n} p_i\vert \psi_i\rangle \langle {\psi}_i\vert
\end{equation}
 and if each $\psi_i$'s are normalised to unity then
 \begin{equation}
    Tr[\hat{\rho}] = \sum_{i=1}^n p_i=1  
 \end{equation}
 In case of mixed state 
 \begin{equation}
   \hat{\rho}^2\neq\hat{\rho} 
 \end{equation}
In Sec 2.4 we have shown that there is an algebraic equivalence between single mode quantized beam of light and one dimensional harmonic oscillator.In other words,if we consider the usual quantization of electromagnetic field, for simplicity restrict ourselves to one mode case only, to describe such fields we can use the harmonic oscillator number states as Fock state for this context.\\
We have the canonical annihilation, creation operators $\hat{a}$ and $\hat{a}^{\dag}$ respectively and they obey the commutation relation $[\hat{a},\hat{a}^{\dag}]=\hat{I}$ and number operator $\hat{N}$ is defined as $\hat{N}=\hat{a}^{\dag}\hat{a}$. Eigen ket corresponding to number operator i.e. Fock state obey the relation
\begin{equation}
\hat{N}\vert n\rangle = \hat{a}^{\dag}\hat{a}\vert n\rangle = n\vert n\rangle
\end{equation}
and $\langle n\vert m\rangle = \delta_{n,m}$. This is the orthonormality condition. The completeness condition is given as
\begin{equation}
    \sum_{n=0}^{\infty} \vert n\rangle\langle n\vert  = \hat{I}
\end{equation}
Matrix elements of $\hat{a}$ and $\hat{a}^{\dag}$ in Fock basis are
\begin{equation}
   \langle n\vert \hat{a} \vert m\rangle = \sqrt{m}\delta_{n,m-1} 
\end{equation}
 and 
 \begin{equation}
  \langle n\vert \hat{a}^{\dag}\vert m\rangle = \sqrt{m+1} \delta_{n,m+1}  
 \end{equation}
The coherent state $\vert z\rangle$ in Fock state representation is 
\begin{equation}
  \vert z\rangle = e^{\frac{-\vert z\vert ^2}{2}}\sum_{n=0}^\infty \frac{z^n}{\sqrt{n!}} \vert n\rangle  
\end{equation}
 where, $n = 0, 1, 2, ...$
and $z \in C $. However, these coherent states are normalized but non-orthogonal, satisfy completeness relation and form an over complete set.\\
Density matrix can be represented in Fock basis as 
\begin{eqnarray}
\hat{\rho}&=& \hat{I}\hat{\rho}\hat{I} \nonumber\\
&=& \left(\sum_{n=0}^{\infty}\vert n\rangle\langle n\vert \right)\hat{\rho}\left(\sum_{m=0}^{\infty}\vert m\rangle\langle m\vert \right) \nonumber\\
&=& \sum_{n=0}^{\infty}\sum_{m=0}^{\infty}{\langle n\vert \hat{\rho}\vert m\rangle}\vert n\rangle\langle m\vert \nonumber\\
&=& \sum_{n=0}^{\infty}\sum_{m=0}^{\infty}\rho_{n,m}|n\rangle\langle m\vert
\end{eqnarray}
In the above we have used the completeness relation using Fock basis. It is not diagonal representation. Using the completeness relation formed out of coherent state basis we are able to express $\hat{\rho}$ as
\begin{eqnarray}
\hat{\rho} &=& \hat{I}\hat{\rho}\hat{I}\nonumber\\
&=& \left(\frac{1}{\pi}\int d^2 {z_1}\vert z_1\rangle \langle z_1\vert \right)\hat{\rho}\left(\frac{1}{\pi}\int d^2 {z_2}\vert z_2\rangle \langle z_2\vert\right)\nonumber\\
&=& \frac{1}{\pi^2}{\int d^2 {z_1}}{\int d^2 {z_2}}\vert z_1\rangle \langle z_2\vert\langle z_1\vert \hat{\rho}\vert z_2\rangle\nonumber
\end{eqnarray}
\begin{eqnarray}
= \frac{1}{\pi^2}{\int d^2 {z_1}}{\int d^2 {z_2}}{\vert z_1\rangle \langle z_2\vert}{e^{-\frac{1}{2}\vert z_1\vert ^2}}{e^{-\frac{1}{2}\vert z_2\vert^2}} \nonumber\\
{\sum_{n=0}^{\infty}\sum_{m=0}^{\infty}}{\frac{{z_1}^{*n}{z_2}^m}
{\sqrt{n!m!}}}\langle n\vert \hat{\rho}\vert m\rangle \nonumber
\end{eqnarray}
\begin{eqnarray}
&=& \frac{1}{\pi^2}\int d^2{z_1}\int d^2{z_2} \vert z_1 \rangle \langle z_2\vert e^{-\frac{1}{2}\vert z_1\vert ^2} e^{-\frac{1}{2}\vert z_2\vert ^2} R\left(z_1^{*},z_{2}\right)\nonumber\\
\end{eqnarray}
where, 
\begin{equation}
   R\left(z_1^{*},z_{2}\right) = \sum_{n=0}^{\infty} \sum_{m=0}^{\infty} \frac{z_{1}^{*n}z_{2}^{m}}
{\sqrt{n!m!}} \langle n\vert \hat{\rho}\vert m\rangle 
\end{equation}
This is the standard way of representing density operator using given basis. \\
R. J. Glauber described in the section VI of his paper published in Sept, 1963[8] this as \emph{...the density operator $\hat{\rho}$ be represented in a unique way by means of a function of two complex variables,
$R\left(z_{1}^{*},z_{2}\right)$ (R. J. Glauber originally used $\alpha^{*},\beta$ as complex variables) which is analytic throughout the finite $z_{1}^{*}$ and $z_{2}$ planes.....}\\
The function $e^{-\frac{1}{2}\vert z_1\vert^2} e^{-\frac{1}{2}\vert z_2\vert ^2} R\left(z_{1}^{*},z_{2}\right)$ in our view is doubly non unique.\\
At this point E. C. G. Sudarshan noted that these coherent state basis are not only complete but over complete. Using over completeness property of coherent states, Sudarshan obtained his famous diagonal coherent state representation of the density matrix.\\
The coherent state $\vert z\rangle$ in Fock state representation is 
\begin{equation}
   \vert z\rangle = e^{\frac{-\vert z\vert^2}{2}} \sum_{n=0}^{\infty} \frac{z^n}{\sqrt{n!}}\vert n\rangle 
\end{equation}
Multiplying both sides of the above equation by $e^{\frac{\vert z \vert^2}{2}}$ we get
\begin{equation}
    e^{\frac{{\vert z\vert}^2}{2}}\vert z\rangle = \sum_{n=0}^{\infty} \frac{z^n}{\sqrt{n!}}\vert n\rangle 
\end{equation}
Taking outer product with $\langle z\vert$ in both sides we find 
\begin{eqnarray}
e^{\frac{{|z|}^2}{2}}\vert z\rangle\langle z\vert &=& \sum_{n=0}^{\infty} \frac{z^n}{\sqrt{n!}} \vert n\rangle \langle z\vert \nonumber\\
or, e^{\frac{|z|^2}{2}} \vert z\rangle\langle z\vert &=& \left(\sum_{n=0}^{\infty} \frac{z^n}{\sqrt{n!}}\vert n\rangle\right)\left(e^{\frac{-|z|^2}{2}} \sum_{m=0}^{\infty} \frac{z^{*m}}{\sqrt{m!}}\vert m\rangle\right)\nonumber\\
or, e^{|z|^{2}} \vert z\rangle\langle z\vert &=& \sum_{n=0}^{\infty} \sum_{m=0}^{\infty} \frac{z^{n} z^{*m}}{\sqrt{n!m!}} \vert n\rangle \langle m\vert \end{eqnarray}
Now we have, $z= re^{i\theta}$, $z^{*} = re^{-i\theta}$\\
Therefore,
\begin{equation}
    e^{r^2} \vert re^{i\theta}\rangle\langle re^{i\theta}\vert = \sum_{n=0}^{\infty} \sum_{m=0}^{\infty} \frac{r^{n+m}}{\sqrt{n!m!}} e^{i(n-m)\theta} \vert n\rangle \langle m\vert
\end{equation}
Multiply both sides of the above equation by $e^{i l\theta}$ and then integrate both sides over $\theta$ in the range $0 < \theta \leqslant 2\pi$.
\begin{eqnarray}
\int_{\theta = 0}^{2\pi} d\theta e^{r^2} e^{il\theta} \vert re^{i\theta} \rangle\langle re^{i\theta}\vert  \nonumber
\end{eqnarray}
\begin{eqnarray}
&=& \sum_{n=0}^{\infty} \sum_{m=0}^{\infty} \frac{r^{n+m}}{\sqrt{n!m!}} \vert n\rangle \langle m\vert\int_{\theta=0}^{2\pi} d\theta {e^{i(l+n-m)\theta}}\nonumber\\
&=& \sum_{n=0}^{\infty} \sum_{m=0}^{\infty} \frac{r^{n+m}}{\sqrt{n!m!}} \vert n\rangle \langle m\vert
2\pi\delta(l+n-m)\nonumber\\
&=& \sum_{n=0}^{\infty} \sum_{m=0}^{\infty} \frac{r^{n+m}}{\sqrt{n!m!}} \vert n\rangle \langle m\vert
\delta(l+n-m)
\end{eqnarray}
Right hand side takes non zero values only when $l = m-n$. Now operate $\frac{\partial}{\partial r}$ for $p$ times on both sides we find 
\begin{align}
\left(\frac{\partial}{\partial r}\right)^{p} \int_{\theta=0}^{2\pi} \frac {d\theta}{2\pi} e^{r^2} e^{il\theta} \vert re^{i\theta}\rangle\langle re^{i\theta}\vert = \nonumber\\
    \sum_{n,m=0}^{\infty} \frac{r^{n+m-p} (m+n)!}{\sqrt{n!m!}(m+n-p)!} \vert n\rangle \langle m\vert \delta(l+n-m)
\end{align}
If we evaluate this equation at $r = 0$, it is easily seen that, in the right hand side only the term with $(m+n-p) = 0$ survives. Thus,
\begin{align}
\left(\frac{\partial}{\partial r}\right)^{p} \int_{\theta=0}^{2\pi} {\frac {d\theta}{2\pi}} e^{r^2}e^{il\theta}\vert re^{i\theta}\rangle\langle re^{i\theta}\vert\vert_{r=0} = \hspace{2cm} \nonumber \\ 
\sum_{n,m=0}^{\infty}\frac{(m+n)!}{(m+n-p)!\sqrt{n!m!}}
\vert n\rangle \langle m\vert
\delta(l+n-m)\delta(m+n-p)  
\end{align}
The right hand side of the above equation gives non zero result
only when the factor in the denominator $(m+n-p)!$ obeys the condition $m+n\geqslant p$ .\\
Therefore, 
\begin{align}
    \left(\frac{\partial}{\partial r}\right)^{m+n} \int_{\theta=0}^{2\pi} \frac {d\theta}{2\pi} e^{r^2} e^{i(m-n)\theta} \vert re^{i\theta}\rangle \langle re^{i\theta}\vert \vert _{r=0}= 
\frac{(m+n)!}{\sqrt{n!m!}} \vert n\rangle \langle m\vert
\end{align}
\begin{align}
    \therefore \vert n\rangle \langle m\vert  = 
    \frac{\sqrt{n!m!}}{(m+n)!} \left(\frac{\partial}{\partial r}\right)^{m+n} \int_{\theta=0}^{2\pi} \frac {d\theta}{2\pi} e^{r^2}e^{i(m-n)\theta} \vert re^{i\theta}\rangle\langle re^{i\theta}\vert \vert_{r=0}
\end{align}
The density matrix in Fock basis is 
\begin{eqnarray}
\hat{\rho}&=& \hat{I}\hat{\rho}\hat{I}\nonumber\\
&=& \left(\sum_{n=0}^{\infty}\vert n\rangle\langle n\vert \right)\hat{\rho}\left(\sum_{m=0}^{\infty}\vert m\rangle\langle m\vert\right)\nonumber\\
&=& \sum_{n=0}^{\infty}\sum_{m=0}^{\infty}{\langle n\vert \hat{\rho}\vert m\rangle}\vert n\rangle\langle m\vert\nonumber\\
&=& \sum_{n=0}^{\infty}\sum_{m=0}^{\infty}\rho_{n,m}\vert n\rangle\langle m\vert 
\end{eqnarray}
This equation infers that any operator written down in Fock basis can be represented also by coherent state basis. But unique feature is that in the coherent state basis we only need diagonal elements. Let's see how this is possible.
\begin{align}
Tr\left[\hat{O} \vert n\rangle \langle m\vert\right] = \frac{\sqrt{n!m!}}{(m+n)!} \left(\frac{\partial}{\partial r}\right)^{m+n} \int_{\theta=0}^{2\pi} \frac {d\theta}{2\pi} e^{r^2}e^{i(m-n)\theta} Tr\left[\hat{O}\vert re^{i\theta}\rangle\langle re^{i\theta}\vert \right]\vert_{r=0} \nonumber\\
\end{align}
\begin{align}
or, \langle m \vert\hat{O} \vert n \rangle =
\frac{\sqrt{n!m!}}{(m+n)!} \left(\frac{\partial}{\partial r}\right)^{m+n} \int_{\theta=0}^{2\pi} \frac {d\theta}{2\pi} e^{r^2}e^{i(m-n)\theta}
\langle re^{i\theta}\vert \hat{O}\vert re^{i\theta}\rangle\vert_{r=0}
\end{align}
From the above equation we see that both diagonal and non diagonal matrix elements of $\hat{O}$ expressed in Fock basis can be represented by considering only the diagonal matrix element of $\hat{O}$ in coherent state basis.\\
Now if we substitute $\vert n\rangle \langle m\vert$ (Eq.(4.29)) in the equation of $\hat{\rho}$ (i.e. Eq.(4.30)) then we find
\begin{align}
 \hat{\rho}= \sum_{n=0}^{\infty}\sum_{m=0}^{\infty}\rho_{n,m}\frac{\sqrt{n!m!}}{(m+n)!} 
\left(\frac{\partial}{\partial r}\right)^{m+n} 
\int_{\theta=0}^{2\pi} \frac{d\theta}{2\pi} e^{r^2}e^{i(m-n)\theta}\vert re^{i\theta}\rangle\langle re^{i\theta}\vert \vert_{r=0} \nonumber
\end{align}
\begin{align}
=\sum_{n=0}^{\infty}\sum_{m=0}^{\infty}\rho_{n,m}\frac{\sqrt{n!m!}}{(m+n)!}\int_{\theta=0}^{2\pi} \frac {d\theta}{2\pi} e^{i(m-n)\theta} 
\left(\frac{\partial}{\partial r}\right)^{m+n}
\left(e^{r^2}\vert re^{i\theta}\rangle\langle re^{i\theta}\vert\right)\vert_{r=0}\nonumber
\end{align}
\begin{align}
 = \sum_{n,m=0}^{\infty}\rho_{n,m}\frac{\sqrt{n!m!}}{(m+n)!}\int_{\theta=0}^{2\pi}\frac {d\theta}{2\pi} e^{i(m-n)\theta}  
\left(\frac{\partial}{\partial r}\right)^{m+n}
\left(e^{r^2}\vert z\rangle\langle z\vert \right)\vert_{r=0}
\end{align}
Using the property of Dirac delta function:
\begin{equation}
  \phi^{(n)}(0) = (-1)^{n} \int dx \delta^{(n)}(x)\phi(x)  
\end{equation}
 we obtain,
\begin{eqnarray}
\hat{\rho} = \sum_{n=0,m=0}^{\infty}\rho_{n,m}\frac{\sqrt{n!m!}}{(m+n)!}\int_{\theta=0}^{2\pi} \frac {d\theta}{2\pi} e^{i(m-n)\theta} 
(-1)^{n+m} \int_{r=0}^{\infty}dr\left[\left(\frac{\partial}{\partial r}\right)^{m+n}\delta(r)\right]
\left(e^{r^2}\vert z\rangle\langle z\vert\right)\nonumber\\
\end{eqnarray}
Therefore,
\begin{align}
\hat{\rho} = 
\int_{r=0}^{\infty} r dr\int_{\theta=0}^{2\pi} d\theta \vert z\rangle\langle z\vert \frac{e^{r^2}}{2\pi r}
\sum_{n,m=0}^{\infty} \rho_{n,m} \frac{\sqrt{n!m!}}{(m+n)!} 
(-1)^{n+m} e^{i(m-n)\theta} 
\left[\left(\frac{\partial}{\partial r}\right)^{m+n}\delta(r)\right]
\end{align}
This is the required diagonal representation.
\section{Normally ordered operator and Optical Equivalence Theorem}
\vspace{0.3cm}
In the previous section we discussed diagonal representation of density operator using diagonal coherent state basis. Now we quote from E. C. G. Sudarshan's paper. \emph{.... This form is particularly interesting since if $ \hat{O} = \hat{a}^{\dag p}\hat{a}^{q}$  (in changed symbol) be any normal operator (i.e. all creation operators to the left of all annihilation operators), its expectation value in the statistical state represented by the density matrix in the ``diagonal" form
\begin{equation}
    \hat{\rho} = \int d^2 z \vert z\rangle \langle z\vert\phi(z) 
\end{equation}
is given by 
\begin{equation}
   Tr\left[\hat{\rho} \hat{O}\right] = Tr\left[ \hat{\rho}\hat{a}^{\dag p}\hat{a}^{q}\right] = \int d^2 z \phi(z) z^{*p} z^{q} 
\end{equation}
This is the same as the expectation value of the complex classical function $z^{*p}z^{q}$ for a probability distribution $\phi(z)$ over the complex plane. The demonstration above shows that any statistical state of the quantum mechanical system may be described by a classical probability distribution over a complex plane, provided all operators are written in the normal ordered form....}\\
We conclude from the particular remark given above \emph{.... any statistical state of the quantum mechanical system may be described by a classical probability distribution over a complex plane, provided all operators are written in the normal ordered form...} is that these statement may convey the incorrect attribution of optical equivalence theorem, since for quantum mechanical state, $\phi(z)$ is not a classical probability function but a singular kind of distribution or  $\phi(z)$  may show negativity at some phase space points.\\
This kind of incorrect statement remains in the paper because E.C.G. wrote this paper with an enormous hurry. It would have been better if sentences like these may have been avoided.\\
Again we quote E.C.G's words \emph{... the classical complex representations can be put in one-to-one correspondence with quantum mechanical density matrices.....}\\
The one to one correspondence between $\phi(z)$ and density operator $\hat{\rho}$ therefore enables the various coherence functions to appear in a similar way as in the Wolf-Mandel formulation\cite{Mandel}. The equation $\hat{\rho} = \int d^2 z {\vert z\rangle \langle z \vert}\phi(z)$ may be used to express $\hat{\rho}_{n, n}$ in terms of $\phi(z)$ i.e.
\begin{eqnarray}
\hat{\rho}_{n,n} &=& \langle n\vert \hat{\rho}\vert n\rangle\nonumber\\ 
&=& \langle n\vert \int d^2 z {\vert z\rangle \langle z\vert }\phi(z)\vert n\rangle\nonumber\\
&=&\int\phi(z)\frac{\vert z\vert^{2n}}{n!}e^{-\vert z\vert ^2} d^{2} z 
\end{eqnarray}
It is seen that Mandel's Poisson transform relation
\begin{equation}
    P(n,T) = \int_{0}^{\infty} P(W) \frac{W^{n}}{n!}e^{-W} dW 
\end{equation}
 is reproduced  in Eq.(4.39) and interpret $\vert z \vert^2$ as the time integrated intensity $W$ of the field.\\ E.C.G  also noted that \emph{.....Hermiticity of $\hat{\rho}$ implies that $\phi(z)$ is a ``real" function in the sense that ${\phi}^*(z^*) = \phi(z)$, but not necessarily positive definite...}. However the positivity of $\hat{\rho}$ does not imply positivity of $\phi$ and therefore brings out a clear distinction between classical and quantum ensembles. It may be emphasized that for all fields, the diagonal coherent representation is formally as if it was a classical ensemble, but with the understanding that the ensemble probability distributions could be non positivite. With this non positivity of the diagonal coherent state representation in general, one can deduce specific quantum effects such as anti-bunching of photons, squeezing etc. Glauber's off diagonal representation of $\hat{\rho}$ in terms of off diagonal projector $\vert z_1\rangle \langle z_2\vert$ as in Eq.(4.19) can in no way 
 lead to this similarity or difference between classical and quantum descriptions.
\section{Expression of $\phi(z)$ for single as well as multimode beam of light}
\vspace{0.3cm}
Comparing 
\begin{equation}
    \hat{\rho} = \int d^2 z {\vert z\rangle \langle z\vert}\phi(z)
\end{equation}
with 
\begin{align}
\hat{\rho} = \int_{r=0}^{\infty}  rdr\int_{\theta=0}^{2\pi} d\theta \vert z\rangle \langle z\vert \frac{e^{r^2}}{2\pi r}
\sum_{n,m=0}^{\infty} \rho_{n,m} \frac{\sqrt{n!m!}}{(m+n)!} 
(-1)^{n+m}e^{i(m-n)\theta}
\left(\frac{\partial}{\partial r}\right)^{m+n}\delta(r)
\end{align}
we find that 
\begin{align}
\phi(z) = \frac{e^{r^2}}{2\pi r}
\sum_{n=0}^{\infty}\sum_{m=0}^{\infty}\rho_{n,m}\frac{\sqrt{n!m!}}{(m+n)!}(-1)^{n+m}e^{i(m-n)\theta} 
\left(\frac{\partial}{\partial r}\right)^{m+n}\delta(r)
\end{align}
Up to this E.C.G considered only for a single mode beam. Then he made a comment how to generalize those above considerations for multi-mode beam. \emph{....These considerations generalize in a straightforward manner to an arbitrary(countable) number of degrees of freedom, finite or infinite. The states are now represented by a sequence of complex numbers $\{ z \}$; and the Fock representation basis is labelled by a sequence of non-negative integers $\{ n \}$ and density matrices by functions of two such sequences $\rho_{ (\{ n \}, \{ m \} )} $. Any such state can be put into one-to-one correspondence with classical probability distributions in a sequence of complex variables $\phi(\{ z \})$ such that the expectation value of any normal ordered operator $\hat{O}(\{\hat{a}^{\dag} \}, \{ \hat{a} \})$ is given by
\begin{equation}
    Tr[\hat{O}(\{ \hat{a}^{\dag} \}, \{ \hat{a} \}), \hat{\rho}]
= \prod_{\lambda} \int d^2 z_{\lambda} O(\{ z^{*} \}, \{ z \}) \phi(\{ z \})
\end{equation}
where the ``real" function $\phi(z)$ is given by.....}
\begin{align}
&\phi(\{z\})= \nonumber\\
&\prod_{\lambda} \frac{e^{r_{\lambda}^2}}{2\pi r_{\lambda}} \sum_{n_{\lambda}
=0, m_{\lambda}=0}^{\infty} \rho_{(\{ n_{\lambda} \}, \{ m_{\lambda} \})} \frac{\sqrt{n_{\lambda}! m_{\lambda}!}} {(m_{\lambda} + n_{\lambda})!} 
(-1)^{(n_{\lambda}+ m_{\lambda})} e^{i(m_{\lambda}-n_{\lambda})\theta_{\lambda}} \left(\frac{\partial}{\partial r_{\lambda}}\right)^{(m_{\lambda}+ n_{\lambda})} \delta(r_{\lambda}) \nonumber
\end{align}
\begin{align}
\therefore \phi(\{ z \})
= \prod_{\lambda} \sum_{n_{\lambda}=0,m_{\lambda}=0}^{\infty} \rho_{(\{ n_{\lambda} \}, \{ m_{\lambda} \})} \frac{\sqrt{n_{\lambda}!m_{\lambda}!}}{2\pi r_{\lambda}(m_{\lambda}+n_{\lambda})!} 
e^{r_{\lambda}^2+i(m_{\lambda}-n_{\lambda})\theta_{\lambda}} 
\left( -\frac{\partial}{\partial r_{\lambda}} \right)^{m_{ \lambda} + n_{ \lambda}}  \delta(r_{\lambda}) \nonumber\\
\end{align}
Again E.C.G. reiterated that \emph{....consequently the description of  statistical states of a quantum mechanical system with an arbitrary (countably infinite) number of degrees of freedom is completely equivalent to the description in terms of classical probability distributions in the same (countably infinite) number of complex variables. In particular, the statistical states of the quantized electromagnetic field may be described uniquely by classical complex linear functions on the classical electromagnetic field. This functional will be ``real" reflecting the Hermiticity of the density matrix; and leads in either version to real expectation values for Hermitian (real) dynamical variables.}\\
As an illustration, consider the vacuum state $\vert 0\rangle$ for which the density operator is given by $\hat{\rho} = \vert 0\rangle \langle 0\vert$. To make use of Eq.(4.43), we at first find the matrix element $\rho_{n,m} = \langle n\vert \hat{\rho}\vert m\rangle$
$= \langle n\vert 0\rangle \langle 0\vert m\rangle$. 
Therefore, $\rho_{n,m}=1$, for $n = m = 0 $ and 
$\rho_{n,m} = 0$, for $n\neq m$. Now we put $n = m = 0$ in Eq.(4.43) we get the possible phase space distribution as 
\begin{equation}
    \phi(z) = \frac{e^{r^2}}{2\pi r} \delta (r)
\end{equation}
Here,  we find the phase space distribution for vacuum state is simply Dirac delta function, hence the state is said to be as good as classical state. \\
Next we consider the pure coherent state $\vert z_0\rangle$ for which 
$\hat{\rho} = \vert z_0\rangle \langle z_{0}\vert $.Its corresponding phase space distribution function is $\delta^{(2)}(z-z_0)$
It is not easy to derive $\phi(z)$ for arbitrary state using E.C.G's way  as $\phi(z)$ involves higher order derivatives of delta function. Now we resort to other way.

\section{ Differently ordered operators,Characteristic functions and Probability distributions }
\vspace{0.3cm}
Expectation value of normally ordered operator  $\hat{a}^{\dag p} \hat{a}^{q}$ with respect to a density operator $\hat{\rho}$ is given by 
\begin{equation}
    Tr\left[\hat{\rho}\hat{a}^{\dag p}\hat{a}^{q}\right] =\int d^2 z\phi(z)z^{*p} z^{q} 
\end{equation}
From the above expression, one can obtain the probability distribution $\phi(z)$ by an inversion formula. It is the inverse Fourier transform of the normally ordered characteristic function.\\
In a more general way, we define normally ordered characteristics function as
\begin{eqnarray}
C_N(\alpha) &=& Tr\left[ \hat{\rho}e^{\alpha \hat{a}{\dag}} e^{-\alpha^{*}
\hat{a}} \right] \nonumber\\
&=& Tr\left[\int d^2 z\phi(z)\vert z\rangle \langle z\vert e^{\alpha \hat{a}{\dag}} e^{-\alpha^{*}
\hat{a}}\right] \nonumber\\
&=& \int d^2 z\phi(z)\langle z\vert e^{\alpha \hat{a}{\dag}}e^{-\alpha ^{*}\hat{a}} \vert z\rangle\nonumber\\
&=& \int d^2 z\phi(z) e^{\alpha z^{*} -\alpha^{*} z}
\end{eqnarray}
We see that $C_N(\alpha)$ is the Fourier transform of $\phi(z)$ where $e^{\alpha z^{*} -\alpha^{*} z}$ is the kernel of Fourier transform in complex plane. $\phi(z)$ is the inverse Fourier transform of $C_N(\alpha)$.
\begin{equation}
    \therefore \phi(z) = \frac{1}{\pi^{2}} \int d^2 {\alpha} {e^{\alpha^{*} z -\alpha z^{*} }} C_{N} (\alpha) 
\end{equation}
Other commonly used characteristic functions are Weyl (or symmetric)
ordered characteristic function $C_W (\alpha)$ and anti-normally ordered characteristic function $C_{A} (\alpha)$.
\begin{eqnarray}
C_W(\alpha) &=& Tr\left[\hat{\rho} e^{\alpha \hat{a}{\dag} - \alpha^{*}
\hat{a}}\right] \nonumber\\
&=& Tr\left[ \hat{\rho} e^{\alpha \hat{a}{\dag}} e^{-\alpha^{*} \hat{a}} e^{\frac{-|\alpha|^2}{2} } \right] \nonumber\\
&=& C_{N} (\alpha) e^{\frac{-|\alpha|^2}{2}}
\end{eqnarray}
Similarly,
\begin{eqnarray}
C_A(\alpha) &=& Tr\left[ \hat{\rho}e^{-\alpha^{*} \hat{a}} e^{\alpha \hat{a}^{\dag}} \right] \nonumber\\
&=& Tr\left[ \hat{\rho}e^{-\alpha^{*} \hat{a} + \alpha 
\hat{a}{\dag}}  e^{\frac{-|\alpha|^2}{2}} \right] \nonumber\\
&=& C_{W} (\alpha) e^{-\frac{|\alpha|^2}{2}}
\end{eqnarray}
These, three characteristic functions are related through the relation : 
\begin{equation}
    C_{W}(\alpha) = C_{N} (\alpha) e^{\frac{-|\alpha|^2}{2}} = C_{A} (\alpha) e^{\frac{|\alpha|^2}{2}}
\end{equation}
Instead of the three characteristic functions, we can introduce the $s$ parameterized function of Cahill and Glauber\cite{Cahill}.\\
\begin{equation}
    C(\alpha,s) = Tr\left[ \hat{\rho} e^{\left( \alpha\hat{a}^{\dag} - \alpha^{*} \hat{a} + s\frac{|\alpha|^2}{2} \right)} \right]
\end{equation}
such that,
\begin{eqnarray}
C(\alpha,0) &=& C_w (\alpha)\nonumber\\
C(\alpha,1) &=& C_N (\alpha)\nonumber\\
C(\alpha,-1)&=& C_A (\alpha)
\end{eqnarray}
Now we find the connections between characteristic functions and various quasi-probability distributions.Already we made a connection between Glauber Sudarshan quasi-probability function $\phi(z)$ and normally ordered characteristic function $C_{N} (\alpha)$ which is 
\begin{eqnarray}
C_{N}(\alpha) &=& Tr\left[ \hat{\rho} e^{\alpha \hat{a}{\dag}} e^{-\alpha^{*}
\hat{a}} \right] \nonumber\\
&=& \int d^2 z \phi(z) e^{\alpha z^{*} -\alpha^{*} z}
\end{eqnarray}
We reiterate that $C_{N}(\alpha)$ is the Fourier transform of $\phi(z)$ where,  $e^{\alpha z^{*} -\alpha^{*} z}$ is the kernel of Fourier transform in complex plane. $\phi(z)$ is the inverse Fourier transform of $C_{N}(\alpha)$.
\begin{equation}
   \therefore \phi(z)=\frac{1}{\pi^2}\int d^2 {\alpha} {e^{\alpha ^* z-\alpha z* }} C_N {(\alpha)}  
\end{equation}
Anti-normally ordered characteristic function $C_{A} (\alpha)$ is defined as
\begin{eqnarray}
C_{A}(\alpha) &=& Tr\left[ \hat{\rho} e^{-\alpha^{*} \hat{a}} e^{\alpha \hat{a}^{\dag}} \right] \nonumber\\
&=& Tr\left[ \hat{I} e^{\alpha \hat{a}^{\dag}} \hat{\rho} e^{-\alpha^{*} \hat{a}} \right] \nonumber \\
&=& \frac{1}{\pi} Tr \left[ \int d^2 z \vert z\rangle \langle z \vert e^{\alpha \hat{a}^{\dag}} \hat{\rho} e^{-\alpha^{*} \hat{a}} \right] \nonumber\\
&=& \frac{1}{\pi} \int d^2 z \langle z\vert e^{\alpha\hat{a}^{\dag}} \hat{\rho} e^{-\alpha^{*} \hat{a}} \vert z \rangle \nonumber \\
&=& \frac{1}{\pi} \int d^2 z e^{\alpha z^{*}-\alpha^{*}z} \langle z \vert \hat{\rho} \vert z\rangle \nonumber\\
&=& \int d^2 z e^{\alpha z^{*}-\alpha^{*}z} Q(z)
\end{eqnarray}
where,
\begin{equation}
    Q(z) = \frac{1}{\pi} \langle z\vert \hat{\rho} \vert z\rangle
\end{equation}
Therefore, $ C_{A} (\alpha)$ is just a Fourier transform of $Q(z)$ function.
Inverse Fourier transform of $C_{A} (\alpha)$ gives $Q(z)$.
\begin{equation}
   Q(z) = \frac{1}{\pi^{2}} \int d^2 {\alpha} e^{\alpha^{*} z - \alpha z^{*}} C_{A} (\alpha) 
\end{equation}
$Q(z)$ is called Hussimi Kano function. \\
Weyl (or symmetric) ordered characteristic function is given by\\
\begin{eqnarray}
C_{W}(\alpha) &=& Tr\left[ \hat{\rho} e^{\alpha \hat{a}^{\dag}-\alpha^{*}
\hat{a}} \right] \nonumber\\
&=& Tr \left[ \int d^2 z \phi(z) \vert z\rangle \langle z \vert e^{\alpha \hat{a}^{\dag}} e^{-\alpha^{*} \hat{a}}
 \right] e^{-\frac{\vert \alpha\vert^2}{2}} \nonumber\\
&=& \left[ \int d^2 z \phi(z) e^{\alpha z^{*}- \alpha^{*} z} \right] e^{-\frac{\vert \alpha\vert^2}{2}} \nonumber\\
&=& C_{N} (\alpha) e^{-\frac{\vert \alpha\vert^2}{2}}
\end{eqnarray}
In a similar way, Wigner phase space distribution function $W(z)$ can be defined as the inverse Fourier transform of the Weyl(or symmetric) ordered characteristic function $C_{W} (\alpha)$.
\begin{equation}
W(z) = \frac{1}{\pi^{2}} \int d^2 {\alpha} e^{\alpha^{*} z-\alpha z^{*} } C_{W}(\alpha)   
\end{equation}
In the above we mentioned three types of phase space distribution functions $\phi(z), Q(z)$ and $W(z)$ for the normal, anti-normal and Weyl's (or symmetric) ordering rules respectively.\\
At this point we want to divulge something related to Sudarshan's $\phi(z)$ function. In this regard, one should notice C. L. Mehta's  remark[11]: \emph{..The only source from which Glauber gets his so called `P-representation' is Sudarshan's published work which he copies, changing $\phi$ to $P$ and $z$ to $\alpha$ ...... It is thus inconvertible that the credit for formulating and discovering the diagonal coherent state representation must go to Sudarshan. It is truly ironic that although these facts are readily accessible, which ought to be called `Sudarshan's diagonal coherent state representation' is dubbed `Glauber's P-representation'. Nor is it correct even as compromise move, to call it the `Glauber-Sudarshan representation'.}\\
In the next para, Mehta also said \emph{...Sudarshan's application of the over completeness of coherent states led to a unique expression for $\phi(z)$, the diagonal coherent state representation of the density operator. As such his work is the first and the only formulation of quantum coherence theory.}\\
We term $\phi(z)$ as Sudarshan's probability function as a move to keep the record straight. Interested reader may go through the references[6,7,2,8] to know the sequence of events that happened in the year 1963.\\
For further illustration, we will evaluate  Sudarshan's probability function $\phi(z)$ corresponding to density operator for thermal or chaotic state of the field 
\begin{equation}
    \hat{\rho} = \frac{1}{1+\Bar{n}} \sum_{n=0}^{\infty}
\left( \frac{\Bar{n}}{1+\Bar{n}} \right)^{n} \vert n\rangle \langle n\vert
\end{equation}
using Hussimi Kano $Q(z)$ probability function and characteristic functions $C_{A}(\alpha)$, $C_N(\alpha)$ of anti-normally, normally ordered operator  respectively as these characteristic functions are interlinked through a simple relation Eq.(4.52).\\
The density operator given above is a mixed state with
\begin{equation}
    \Bar{n} = \frac{1}{exp\left( \frac{\hbar\omega}{K_{B} T}\right)-1}.
\end{equation}
where, $\Bar{n}$ is the mean number of photons in a thermal reservoir.\\
First we calculate Hussimi Kano probability function $Q(z)$.
\begin{eqnarray}
Q(z) &=& \frac{1}{\pi} \langle z\vert \hat{\rho}\vert z\rangle \nonumber\\
&=& \frac{1}{\pi} e^{-\vert z\vert^2} \sum_{m} \sum_{n} \langle m\vert \hat{\rho} \vert n\rangle \frac{z^{*m} z^{n}}{\sqrt{m!n!}} \nonumber\\
&=& \frac{e^{-|z|^2}}{\pi(1+ \Bar{n})} \sum_{n} \left( \frac{\Bar{n}}{(1+\Bar{n}} \right)^{n} \frac{ \left( z^{*} z \right)^{n}}{n!} \nonumber\\
&=& \frac{1}{\pi(1+\Bar{n})} exp\left( -\frac{|z|^2}{1+\Bar{n}}\right)
\end{eqnarray}
Now we evaluate $C_{A}(\alpha)$ which is the Fourier transform of $Q(z)$ function.
\begin{eqnarray}
C_{A}(\alpha) &=& \int d^2 z e^{\alpha z^{*}-\alpha^{*} z} Q(z)\nonumber\\
&=& \frac{1}{\pi(1+\Bar{n})} \int d^2 z e^{\alpha z^{*}-\alpha^{*}z } exp\left( -\frac{|z|^2}{1+\Bar{n}} \right) \nonumber\\
\end{eqnarray}
Let, $z = \frac{q+ip}{\sqrt{2}}$, $\alpha = \frac{x + iy}{\sqrt{2}}$, $d^2 z = \frac{dqdp}{2}$
\begin{equation}
C_{A}(x,y) = \frac{1}{2\pi (1 + \Bar{n})} \int {dqdp} \hspace{0.1cm}
e^{i(yq-xp)} exp\left[ -\frac{q^{2} + p^{2}} {2(1+\Bar{n})} \right] 
\end{equation}
Using standard formula
\begin{equation}
\int e^{-as^{2}} e^{\pm \beta s} ds = \sqrt{\frac{\pi}{a}}. e^{\frac{\beta ^{2}}{4a}} 
\end{equation}
we obtain,
\begin{equation}
  C_{A}(\alpha) = exp[-(1+\Bar{n})|\alpha|^2]  
\end{equation}
We know the relation between $C_N(\alpha)$ and $C_A(\alpha)$ as
\begin{equation}
    C_N(\alpha)=C_A(\alpha)exp(|\alpha|^2)=exp(-\Bar{n}|\alpha|^2)
\end{equation}
As Glauber Sudarshan function $\phi(z)$ is the inverse Fourier transform of $C_N(\alpha)$. So, 
\begin{eqnarray}
\phi(z) &=& \frac{1}{\pi^{2}} \int d^2 {\alpha} e^{\alpha^{*}z-\alpha z^{*}} C_{N} (\alpha) \nonumber\\ 
&=& \frac{1}{\pi^{2}} \int d^2 \alpha e^{\alpha^{*} z - \alpha z^{*}} exp(-\Bar{n}|\alpha|^2)\nonumber\\
&=& \frac{1}{\pi\Bar{n}} exp\left(-\frac{|z|^2}{\Bar{n}}\right)
\end{eqnarray}
The Sudarshan's probability function $\phi(z)$ corresponding to density operator for thermal or chaotic state of the field is Gaussian. As it is a true probability distribution not quasi probability function, so thermal or chaotic field state is a classical state according to Sudarshan's criterion.\\
The probability distribution $\phi(z)$ [as in Eq.(4.70)] for thermal or chaotic state of the field is equivalent to the result for $\rho_{n,n}$. \\
Take the matrix elements taking $\hat{\rho}$  in diagonal coherent state basis.
\begin{eqnarray}
\rho_{n,m} &=& \langle n\vert \hat{\rho}\vert m\rangle \nonumber\\
&=& \int d^2 z \phi(z)\langle n \vert z\rangle \langle z \vert m\rangle \nonumber\\
&=& \frac{1}{\sqrt{n!m!}} \int d^2 z \phi(z)e^{-|z|^2} z^{n} z^{*m}
\end{eqnarray}
Now we put $\phi(z)$ [as in Eq.(4.70)] in the above matrix element and find 
\begin{eqnarray}
\rho_{n,m} &=& \frac{1}{\sqrt{n!m!}} \int d^2 z \frac{1}{\pi\Bar{n}} exp(-\frac{|z|^2}{\Bar{n}})
e^{-|z|^2} z^{n}z^{*m} \nonumber\\
&=& \frac{1}{\sqrt{n!m!}} \frac{1}{\pi\Bar{n}} \int d^2 z exp(-\frac{|z|^2}{\Bar{n}})
e^{-|z|^2} z^{n}z^{*m} \nonumber\\
\end{eqnarray}
Let, $z = re^{i\theta}$; $z\in C$
\begin{align}
\rho_{n,m} = \frac{1}{\sqrt{n!m!}} \frac{1}{\pi\Bar{n}}
\int_{0}^{\infty} dr r^{n+m+1} e^{-r^{2}(\frac{1}{\Bar{n}}+1)} . \nonumber\\
\int_{0}^{2\pi} d\theta e^{i(n-m)\theta} \nonumber
\end{align}
\begin{align}
&=& \frac{1}{\sqrt{n!m!}} \frac{1}{\pi\Bar{n}}
\int_{0}^{\infty} dr r^{n+m+1} e^{-Cr^{2}} 2\pi \delta_{n,m} 
\end{align}
where, $C =(\frac{1}{\Bar{n}}+1)$.\\
On evaluating the $\theta$-integral part by using the property of Delta function, we can get the non-zero value only when, $n = m$.
\begin{eqnarray}
\therefore \rho_{n,n} &=& \frac{2\pi}{n!\pi\Bar{n}}
\int_{0}^{\infty} dr r^{2n+1} e^{-Cr^{2}}\nonumber\\
&=& \frac{2\pi}{n!\pi\Bar{n}} \frac{n!}{2C^{n+1}}
\end{eqnarray}
On substituting, $C = \frac{1}{\Bar{n}}+1$ in the above equation we find 
\begin{equation}
\rho_{n,n} = \frac{1}{1+\Bar{n}} \left( \frac{\Bar{n}}{1+\Bar{n}} \right)^{n}
\end{equation}
\textit{What is $\rho_{n,n}$}?\\ 
The number or Fock states form a complete set, hence a general expansion of $\hat{\rho}$ is
$\hat{\rho} = \sum_{n,m=0}^{\infty} \rho_{n,m}\vert n\rangle  \langle m\vert$.
The coefficients $\rho_{n,m}$ are complex and they are infinitely many. All these expansion coefficients are not useful where only the photon number distribution is of interest. Only diagonal terms are important then we use
\begin{equation}
    \hat{\rho} = \sum_{n=0}^{\infty} \rho_{n,n} \vert n\rangle  \langle n\vert
\end{equation}
Here $\rho_{n,n}$ is a probability distribution giving the probability of having $n$ photons in the mode. This is not a general representation for all fields but it is useful for certain fields e.g. a chaotic field or thermal field, which has no phase information.\\
\textit{What Sudarshan said in this context?}\\ 
In his paper\cite{E.C.G} expression for Phase space probability function is given in Eq.(6).
\begin{equation}
    \phi(z) = \frac{e^{r^{2}}}{2\pi r}
\sum_{n=0}^{\infty} \sum_{m=0}^{\infty} \rho_{n,m} \frac{\sqrt{n!m!}}{(m+n)!}
(-1)^{n+m} e^{i(m-n)\theta}
\end{equation}
At this point we quote Sudarshan and his Eq.(7)(see \cite{E.C.G}) : \emph{...Eq.(6) for $\phi(z)$ in terms of $\rho_{n,m}$ can be inverted to yield 
\begin{equation}
   \rho_{n,m} = \frac{1}{\sqrt{n!m!}} \int d^2 z \phi(z)e^{-|z|^2} z^{n} z^{*m}
\end{equation}
If we do this for the Gaussian function, we obtain the Bose-Einstein distribution, diagonal in the occupation number sequences corresponding to the equal weightage of all phase angles. It is worth pointing out that this result reproduces the Purcell-Mandel derivation for photoelectric counting statistics}.\\
For the sake of the reader, we again herewith write  the probability distribution of having $n$ photons i.e. $\rho_{n,n}$ and the Mandal's Poisson transformation relation, $P(n,T)$ respectively, for the better perception
\begin{eqnarray}
\rho_{n,n} &=& \frac{1}{n!} \int d^2 z \phi(z) e^{-|z|^2} z^{2n}\\ 
 P(n,T) &=& \frac{1}{n!} \int_{0}^{\infty} dW P(W) e^{-W} W^{n}
\end{eqnarray}
\subsection{Determination of $\phi(z)$ using Anti-Diagonal method}
\vspace{0.3cm}
The anti-diagonal method of C. L. Mehta\cite{Mehta1967} to evaluate $\phi(z)$ is particularly useful from a practical point of view. We previously mentioned that the weight factor $\phi(z)$ is the inverse Fourier transform of characteristic function $C_N(\alpha)$ and in the following sections we will show that $\phi(z)$ can be regarded as the limit of a sequence of tempered distributions for some specific cases. In this sense it is a well defined generalized function. E.C.G's original explicit expression for $\phi(z)$ is a formal series expansion involving derivatives of Dirac delta function and it is hard to use. Mehta derived a simple expression for $\phi$ which is a well behaved function.
Here we express density operator $\hat{\rho}$ as  
\begin{equation}
   \hat{\rho} = \int \phi(z) {\vert z\rangle \langle z\vert}d^2 z 
\end{equation}
We multiply the above equation by $\langle -u \vert $ on the left and $\vert u\rangle$ on the right. 
where, both $\vert -u\rangle$ and $\vert u\rangle$ are coherent states.
\begin{eqnarray}
\langle -u\vert \hat{\rho}\vert u\rangle &=& \int \phi(z) {\langle -u \vert z\rangle \langle z|u\rangle}d^2 z \nonumber\\
&=& e^{-|u|^2}\int \phi(z)e^{-|z|^2}e^{z^{*} u - zu^{*}}d^2 z \nonumber\\
&=& e^{-|u|^2}g(u) 
\end{eqnarray}
Now from the above equation, we see that,
\begin{equation}
  g(u) = e^{|u|^2} \langle -u \vert \hat{\rho}\vert u\rangle 
\end{equation}
where, 
\begin{equation}
     g(u) = \int \phi(z)e^{-|z|^2}e^{z^{*} u-zu^{*}}d^2 z 
\end{equation}
The above equation for $g(u)$ can be seen as a Fourier transform of $\phi(z)e^{-|z|^2}$ in a complex plane. Taking the inverse Fourier transform of $g(u)$ one can find $\phi(z)e^{-|z|^2}$ i.e.
\begin{eqnarray}
\phi(z)e^{-|z|^2} &=& \frac{1}{\pi^2}\int e^{|u|^2}\langle -u|\hat{\rho}|u\rangle e^{zu^*-z^*u}d^2 u \nonumber\\
or, \phi(z) &=& \frac{e^{|z|^2}}{\pi^2}\int e^{|u|^2}\langle -u|\hat{\rho}|u\rangle e^{zu^*-z^*u}d^2 u \nonumber\\
\end{eqnarray}
The matrix element $\langle -u\vert \hat{\rho}\vert u\rangle$ should decrease with $u$ for large $u$ more rapidly than $e^{-\vert u\vert^2}$, otherwise the integral will not exist in the ordinary sense of function theory. So care must be taken in regard to the convergence of the integral. \\
As an illustration of Anti Diagonal method, we consider a quantum mechanical state for a single mode stabilised coherent laser light  $\vert \beta\rangle$ for which density operator is given by 
\begin{equation}
  \hat{\rho}=\vert \beta\rangle \langle \beta \vert   
\end{equation}
Making use of  this $\rho$, we first obtain the matrix element $\langle -u\vert \rho \vert u \rangle$ and then on substituting it in Eq.(4.84) we get the expression of $\phi(z)$ as follow:
\begin{eqnarray}
\langle -u\vert \rho \vert u \rangle &=& 
\langle -u\vert \beta\rangle \langle \beta\vert u\rangle \nonumber\\ &=& e^{-|\beta|^2} e^{-|u|^2} e^{-u^{*} \beta + \beta^{*}u} 
\end{eqnarray}
Now,
\begin{eqnarray}
\phi(z) &=& \frac{e^{|z|^2}}{\pi^2} \int e^{|u|^2} \langle -u\vert \beta\rangle \langle \beta\vert u\rangle e^{zu^{*}-z^{*}u} d^2 u  \nonumber\\
&=& \frac{e^{|z|^2}}{\pi^2}e^{-|\beta|^2}\int {e^{u^* (z-\beta)}e^{-u(z^*-\beta^*)} d^2u} \nonumber\\
&=& {e^{|z|^2}} e^{-|\beta|^2} \delta^{(2)} (z-\beta) \nonumber\\
&=&  {\delta^{(2)}}(z-\beta)
\end{eqnarray}
Therefore, $\phi(z)$ corresponding to a stabilised laser light is Dirac delta function in a complex plane. Hence according to Sudarshan's criterion stabilised laser light is a classical light.\\
Now we take $\hat{\rho} = |n\rangle \langle n|$ for $n$ photon state. First we find 
\begin{equation}
    \langle -u \vert \hat{\rho} \vert u\rangle = \langle -u\vert n\rangle \langle n\vert u\rangle = e^{-|u|^2} \frac{(-{u}^*u)^n}{n!}
\end{equation}
Thus,
\begin{eqnarray}
\phi(z) &=& \frac{e^{|z|^2}}{n!\pi^2} \int (-{u}^{*}u)^{n} e^{zu^{*}-z^{*}u} d^2 u\nonumber\\
&=& \frac{e^{|z|^2}}{n!} \frac{\partial^{2n}}{\partial {z^{n}} \partial {z^{*n}}} \delta ^{(2)} (z,z^{*})
\end{eqnarray}
So the phase space distribution corresponding to $\vert n\rangle$ state is $2n$ order derivatives of Dirac delta function, called a tempered distribution. It is more singular than delta function. It is not a true probability function but a quasi-probability function. This kind of distribution is a signature of non classical state according to E. C. G. Sudarshan.
\subsection{More on Sudarshan's $\phi(z)$ phase space distribution}
\vspace{0.3cm}
We already know that for a given density operator $\hat{\rho}$ it can be represented in diagonal coherent state basis i.e.
\begin{equation*}
    \hat{\rho} = \int d^2 z {\vert z\rangle \langle
z\vert}\phi(z)
\end{equation*}
where $\phi(z)$ is a phase space distribution function or a probability distribution. E.C.G. mentioned that \emph{...$\phi(z)$ not necessarily positive definite.}\\
If $\phi(z)$ has the properties of a classical probability distribution, the state $\hat{\rho}$ is a statistical mixture of coherent states,hence it is called classical. For classical $\hat{\rho}$, $\phi(z)$ can be interpreted as a true probability distribution over the complex plane. Thermal or chaotic state of the field is a classical state. Its  $\phi(z)$ is Gaussian which is a true probability distribution. For stabilised laser emitted coherent light $|z_0\rangle$, the phase space distribution $\phi(z)$ is just Dirac delta function in complex plane.The state is called classical according to the criterion given by E.C.G.\\
Conversely, a state is referred to as non classical if  $\phi(z)$ shows some negativities or it has singularities more than Dirac delta function.
For the single photon state $\vert 1\rangle$ and  $n$ photon state $\vert n\rangle$ is said to be non classical state as their $\phi(z)$ is more singular or highly singular than Dirac delta function.\\
A state is referred to also as non-classical if the $\phi(z)$ shows some negativities at few phase space points over the complex plane. Example of such states are Single-photon added thermal states (SPATS),  Single photon added Coherent state (SPACS). For these states $\phi(z)$ has significant negativities, so it can be termed as quasi-probability. This kind of signature suggests that the corresponding state is non classical.\\
It is interesting  to see how classical states like thermal state,coherent state  turn non classical, when a single quantum of radiation(photon) excites them.\\
We already know that the phase space distribution $\phi(z)$ corresponding to the thermal or chaotic state of field is Gaussian which is a true probability function.\\
Density matrix of single mode thermal state of system in thermal equilibrium, is given by 
\begin{equation}
  \hat{\rho}_{th} = \frac{1}{1+\Bar{n}} \sum_{n=0}^{\infty}
\left( \frac{\Bar{n}}{1+\Bar{n}}\right)^{n} \vert n\rangle \langle n\vert  
\end{equation}
The density operator given above is a mixed state with
\begin{equation}
\Bar{n} = \frac{1}{exp\left(\frac{\hbar\omega}{K_{B} T}\right)-1}
\end{equation}
where, $\Bar{n}$ is the mean number of photons in a thermal reservoir.\\
Using characteristic functions of differently ordered operators we evaluated in sec.4.5 that the Glauber Sudarshan probability function corresponding to the thermal state is 
\begin{equation}
   \phi(z) = \frac{1}{\pi\Bar{n}} exp\left(-\frac{|z|^2}{\Bar{n}}\right) 
\end{equation}
Single Photon added thermal state (SPATS) is obtained through the application of creation operator on the thermal state i.e.
\begin{equation}
    \hat{\rho}_{SPATS} = \hat{a}^{\dag}\hat{\rho}_{th}\hat{a}
\end{equation}
We know that
\begin{eqnarray}
    \hat{\rho}_{th} &=& \frac{1}{1+\Bar{n}} \sum_{n=0}^{\infty}
(\frac{\Bar{n}}{1+\Bar{n}})^n |n\rangle \langle n \vert \nonumber\\
&=& A\sum_{n=0}^{\infty} x^{n} \vert n\rangle \langle n\vert
\end{eqnarray}
where, $A = \frac{1}{1+\Bar{n}} = 1-x$ and $x = e^{-\frac{\hbar\omega}{k_{B} T}} $; with $0\leqslant x\leqslant 1$.
\begin{eqnarray}
\hat{\rho}_{SPATS} &=& \hat{a}^{\dag} \hat{\rho}_{th} \hat{a} \nonumber\\
&=& (1-x)\sum_{n=0}^{\infty} x^{n}  \hat{a}^{\dag} \vert n\rangle \langle n\vert \hat{a} \nonumber\\
&=& (1-x)\sum_{n=0}^{\infty} x^{n}  (n+1)\vert n+1\rangle \langle n+1 \vert \nonumber\\
&=& (1-x)\sum_{m=0}^{\infty} m x^{m-1} \vert m\rangle \langle m\vert 
\nonumber\\
&=& (1-x) \frac{\partial}{\partial x} \left( \frac{1}{1-x} \hat{\rho}_{th} \right)
\end{eqnarray}
Now, we find Normalized photon added thermal state(SPATS). Let, $C$ be the normalization constant.
\begin{equation}
Tr[C\hat{\rho}_{SPATS}]=C(1-x)\frac{\partial}{\partial x}\large(\frac{1}{1- x}Tr[\hat{\rho}_{Th}]\large)
\end{equation}
As $Tr[C\hat{\rho}_{SPATS}]=1$, it implies that 
\begin{equation}
C(1-x)\frac{\partial}{\partial x}\large(\frac{1}{1- x}\large)=1
\end{equation}
 where, $Tr[\hat{\rho}_{Th}]=1$
Therefore, $C=(1-x)$.
Normalized photon added thermal state(SPATS) is given by
\begin{equation}
   \hat{\rho}_{SPATS} = (1-x)^2\frac{\partial}{\partial x} \left( \frac{1}{1-x}{\hat{\rho_{th}}} \right) 
\end{equation}
The above operator equation implies that 
\begin{eqnarray}
    \phi_{SPATS}(z) &=& (1-x)^2 \frac{\partial}{\partial x} \left( \frac{1}{1-x} \phi_{th}(z) \right) \nonumber\\
&=& (1-x)^2 \frac{\partial}{\partial x} \left( \frac{1}{1-x} \frac{1}{\pi \Bar{n}} e^{-\frac{|z|^2}{\Bar{n}}} \right) \nonumber\\
\end{eqnarray}
On substituting, $\Bar{n} = \frac{x}{1-x}$, the above expression becomes
\begin{eqnarray}
\phi_{SPATS}(z)&=& \frac{e^{|z|^2}}{\pi}(1-x)^2\frac{\partial}{\partial x}\left(\frac{1}{x}e^{-\frac{|z|^2}{x}}\right) \nonumber\\
&=& \frac{e^{|z|^2}}{\pi}(1-x)^2{e^{-\frac{|z|^2}{x}}}\frac{1}{x^3}\left(|z|^2 -x\right) \nonumber\\
\end{eqnarray}
GS probability function in terms of $\Bar{n}$ is as follows
\begin{equation}
   \phi_{SPATS}(z) = \frac{1}{\pi{\Bar{n}^3}}\left[(\Bar{n}+1)|z|^2 -\Bar{n}\right]exp\left(-\frac{|z|^2}{\Bar{n}} \right) 
\end{equation}
The GS probability function given above is negative for $z = 0$ for any $\Bar{n}>0$. It represents a so called SPATS i.e. a single photon on a thermal background. This example justifies that a rigorous analysis of the singularities of the GS distributions is indispensable for a profound understanding of the non classical features of mixed quantum states of light. SPATS is non classical according to Sudarshan' criterion and it justifies his remark  \emph{...$\phi(z)$ not necessarily positive definite} given in paper\cite{E.C.G}. \\
We know that Coherent states are the analogs of classical radiation fields according to Sudarshan's criterion. Its density operator given by
$\hat{\rho} = \vert z_0\rangle \langle z_0\vert$ corresponds to GS distribution $\phi(z) = \delta^{(2)}(z-z_0)$. It is interesting to see how these states turn non  classical, when a single quantum of radiation excites them. This non classical state is called single photon added coherent states(SPACS).  SPACS are obtained by application of creation operator $\hat{a}{\dag}$ on a coherent state $\vert z_0\rangle$. Normalized SPACS is given by 
\begin{eqnarray}
\vert SPACS\rangle &=& \frac{\hat{a}^{\dag}|z_0\rangle}{(1+|z_0|^2)} \\
\hat{\rho}_{SPACS} &=& \frac{1}{\left(1 + |z_0|^{2}\right)^2} \hat{a}^{\dag} \vert z_0\rangle\langle z_0\vert \hat{a}
\end{eqnarray}
Now we would like to find its corresponding Sudarshan's probability distribution.
\begin{equation}
    \hat{\rho}_{SPACS} = C\hat{a}^{\dag}\vert z_0\rangle\langle z_0\vert\hat{a}
\end{equation}
where, $C = \frac{1}{\left( 1 + |z_0|^2 \right)^{2}}$\\
Now, we show that 
\begin{equation}
    \hat{a}^{\dag} \vert z_{0} \rangle \langle z_{0} \vert = \left( z_{0}^{*} + \frac{\partial}{\partial z_{0}} \right) \vert z_{0} \rangle \langle z_{0} \vert 
\end{equation}
We take, 
\begin{eqnarray}
\frac{\partial}{\partial z_{0}} \vert z_{0} \rangle \langle z_{0} \vert 
&=& \frac{\partial}{\partial z_{0}} \left( e^{-\vert z_{0}\vert^{2}} e^{z_{0} \hat{a}^{\dag}} \vert 0 \rangle \langle 0 \vert e^{z_{0}^{*} \hat{a}} \right) \nonumber 
\end{eqnarray}
\begin{align}
 \hspace{0.5cm} = -z_{0}^{*} e^{-\vert z_{0}\vert^{2}} e^{z_{0}
\hat{a}^{\dag}} \vert 0 \rangle \langle 0 \vert  e^{z_{0}^{*} \hat{a}} + \nonumber\\
e^{-\vert z_{0}\vert^{2}} \hat{a}^{\dag} e^{z_{0}\hat{a}^{\dag}} \vert 0 \rangle \langle 0 \vert e^{z_{0}^{*}\hat{a}} \nonumber
\end{align}
\begin{eqnarray}
&=& -z_{0}^{*} \vert z_{0} \rangle \langle z_{0} \vert + \hat{a}^{\dag} \vert z_{0} \rangle \langle z_{0} \vert \nonumber\\
&=& \left( \hat{a}^{\dag} - z_{0}^{*} \right) \vert z_{0} \rangle \langle z_{0} \vert
\end{eqnarray} 
\begin{equation}
    \therefore \hat{a}^{\dag} \vert z_{0} \rangle \langle z_{0} \vert =\left( z_{0}^{*} + \frac{\partial}{\partial z_{0}} \right) \vert z_{0}\rangle \langle z_{0}\vert
\end{equation}
In a similar way one can obtain,
\begin{eqnarray}
\frac{\partial}{\partial z_{0}^{*}}  \vert z_{0} \rangle \langle z_{0} \vert    &=& \vert z_{0} \rangle \langle z_{0} \vert (\hat{a} - z_{0}) \\
or, \left( z_{0} + \frac{\partial}{\partial z_{0}^{*}} \right) 
\vert z_{0} \rangle \langle z_{0} \vert  &=& \vert z_{0} \rangle \langle z_{0} \vert \hat{a}
\end{eqnarray}
Again we write, 
\begin{eqnarray}
\hat{\rho}_{SPACS} &=& C \hat{a}^{\dag} \vert z_{0} \rangle \langle z_{0} \vert \hat{a} \nonumber \\
 &=& C \left( z_{0}^{*} + \frac{\partial}{\partial z_{0}} \right) \vert z_{0} \rangle \langle z_{0} \vert  \hat{a} \nonumber \\
&=& C \left( z_{0}^{*} + \frac{\partial}{\partial z_{0}} \right) \left( z_{0} + \frac{\partial}{\partial z_{0}^{*}} \right) \vert z_{0} \rangle \langle z_{0}\vert  \nonumber
\end{eqnarray}
\begin{align}
\hat{\rho}_{SPACS} = 
C [ \vert z_{0} \vert^{2} \vert z_{0} \rangle \langle z_{0} \vert +  z_{0}^{*} \frac{\partial}{\partial z_{0}^{*}} \vert z_{0} \rangle \langle z_{0}\vert +  
\frac{\partial}{\partial z_{0}} \left( z_{0} \vert z_{0} \rangle \langle z_{0}\vert \right) 
+ \frac{\partial^{2}}{\partial z_{0} \partial z_{0}^{*}} \vert z_{0} \rangle \langle z_{0} \vert]
\end{align}
Sudarshan's $\phi(z)$ function of a coherent state has been identified as         $\delta^{(2)}(z-z_0)$ .\\
So, Sudarshan's $\phi_{SPACS}(z)$ is
\begin{align}
    \phi_{SPACS}(z) =
    C \left[ \left(1 + |z_{0}|^2\right) + z_{0}^{*}\frac{\partial}{\partial z_{0}^{*}}  + z_{0} \frac{\partial}{\partial z_{0}}  + \frac{\partial ^{2}}{\partial z_{0} \partial z_{0}^{*}}  \right] \delta^{(2)}(z-z_{0})   
\end{align}
Here, we see that $\phi_{SPACS}(z)$ is a highly singular function because it comprises of first and second order derivatives of Dirac delta function hence the state $\hat{\rho}_{SPACS}$ is a non classical state.\\
According to Sudarshan's criterion, vacuum state $\vert 0 \rangle$ is as good as classical because Sudarshan's Probability function $\phi(z)$ takes the form of a  delta function in complex plane i.e. $\delta^{(2)}(z)$. If squeezing operator acts on it then a squeezed state is formed i.e. $\vert \xi \rangle = exp\left[ \frac{1}{2} \left( \xi \hat{a}^{\dag 2}-\xi^{*} \hat{a}^{2} \right)\right] \vert 0 \rangle$. One can show that this state corresponds to negative $\phi(z)$ over some phase space points and hence this is a state having no classical analogue.\\
\chapter{$\phi(z)$ and evidence for the quantum nature of light}
\vspace{0.3cm}
The experiments of classical optics involve first order correlation function e.g. amplitude amplitude correlation. Interference using single photon laser beam bears the signature of it. Most of the experiments of classical optics may be categorised as being in the domain of one photon or linear optics.\\
The advent of photon correlation experiments, as pioneered by Hanbury Brown and Twiss\cite{HBT}  began a fundamentally different era of nonlinear optics. The quantum mechanical interpretation of the Hanbury Twiss effect and other photon correlation experiments were given by R. J. Glauber[6]. He invoked a classical description of a fluctuating electromagnetic field. In the same paper Glauber pointed out that photon correlation experiments offer the possibility of observing a uniquely quantum-mechanical effect -namely photon anti-bunching. \\
This anti-bunching effect is a result of anti-correlation. But how one can explain this quantum mechanical phenomena using classical probability which is always being positive definite. To answer this question Sudarshan's $\phi(z)$ representation provides us that \emph{...$\phi(z)$ is a `real' function in the sense that $\phi^{*}(z^{*}) = \phi(z)$, but not necessarily positive definite.} This is a remarkable statement by Sudarshan to interpret the nonclassicality.\\
The first experiment beyond one photon optics was performed in 1956 by Hanbury Brown Twiss\cite{HBT}. This is an intensity correlation experiment and it is a measurement of $\langle I(t+ \tau)I(t)\rangle$. In essence this experiment measures the joint photo-count probability of detecting the arrival of a photon at time $t$ and another photon at $t+\tau$ Second order correlation function of the radiation field introduced by Glauber may describe the measurements of the correlation phenomena. Normalized Second order coherence function is defined as 
\begin{equation}
    g^{(2)} (\tau) = \frac{\langle \hat{E}^{(-)}(t)\hat{E}^{(-)}(t+\tau)\hat{E}^{(+)}(t+\tau)\hat{E}^{(+)}(t)\rangle}{\langle \hat{E}^{(-)}(t)\hat{E}^{(+)}(t)\rangle \langle\hat{E}^{(-)}(t+\tau)\hat{E}^{(+)}(t+\tau)\rangle}
\end{equation}
where, $\hat{E}^{(+)}(t)$ and $\hat{E}^{(-)}(t)$ are the positive and negative frequency parts of the electromagnetic field operator respectively.\\
Now we attempt to give an interpretation of $g^{(2)}(0)$ for a single mode electromagnetic field.
\begin{eqnarray}
g^{(2)} (0) &=& \frac{\langle \hat{a}^{\dag}\hat{a}^{\dag}\hat{a}\hat{a}\rangle}{{\langle \hat{a}^{\dag}\hat{a}\rangle}^2}\nonumber\\
&=& \frac{\langle \hat{N}(\hat{N}-I)\rangle}{{\langle \hat{N}\rangle}^2}\nonumber\\
&=& \frac{{\langle \hat{N}\rangle}^2 +{\langle \hat{N}^2\rangle}-{\langle \hat{N}\rangle}-{\langle \hat{N}\rangle}^2}{{\langle \hat{N}\rangle}^2}\nonumber\\
&=& 1+\frac{{\langle \hat{N}^2\rangle}-{\langle \hat{N}\rangle}-{\langle \hat{N}\rangle}^2}{{\langle \hat{N}\rangle}^2}\nonumber\\
&=& 1+ \frac{(\Delta N)^2-{\langle \hat{N}\rangle}}{{\langle \hat{N}\rangle}^2}
\end{eqnarray}
where, ${\langle \hat{N}\rangle}$ and $(\Delta N)^2$ are the mean and variance of the photon number distribution.\\
Thermal light has a power law photon number distribution with variance $(\Delta N)^2 ={\langle \hat{N}\rangle}^2+{\langle \hat{N}\rangle}$ which gives $g^{(2)} (0)=2$. \\
Coherent light from a highly stabilised laser has a poissonian photon number distribution with $(\Delta N)^2 = {\langle \hat{N}\rangle}$ leading to $g^{(2)} (0)=1$.  \\
Electromagnetic field which has photon number distribution narrower than poissonian  $(\Delta N)^2 <{\langle \hat{N}\rangle}$ then we obtain $g^{(2)} (0)< 1$. This is just the opposite effect to the photon bunching observed for thermal light; call it photon antibunching. In this case photon number fluctuations get reduced to an amount which is below that of a poissonian distribution.\\
 We describe electromagnetic field using classical theory using a fluctuating complex field amplitude $E$. These fluctuations are  taken into account by introducing a probability distribution
 $P(E)$. The normalized second order coherence function $g^{(2)} (0)$ looks like 
 \begin{equation}
     g^{(2)} (0)=1+\frac{\int P(E)\left(|E|^2 -\langle |E|^2\rangle\right)^2 d^2 E}{(\langle |E|^2\rangle)^2}
 \end{equation}
 For  thermal or chaotic field; $P(E)$ is Gaussian probability distribution then $g^{(2)} (0)=2$. The distribution $P(E)$ for stabilized laser beam is a delta function and hence $g^{(2)} (0)=1$.\\
 Hence, the Hanbury Brown Twiss effect for thermal or chaotic radiation and coherent laser field is well described by classical theory.Classical optical coherence theory shows that $g^{(2)} (0)\geq 1$. It does not allow photon antibunching for which $g^{(2)} (0)< 1$. At this point Sudarshan's diagonal coherent state representation of density operator of the field resort us in  such a way that formal appearance of $g^{(2)} (0)$ using $\phi(z)$ looks like the expression (5.3) that appeared in Classical theory.
 
 The expression of $g^{(2)}(0)$ given in eq.(5.2) reads

$g^{(2)} (0)=1+{\frac{(\Delta N)^2-{\langle \hat{N}\rangle}}{{\langle \hat{N}\rangle}^2}}$.Calculate $(\Delta N)^2,\langle \hat{N}\rangle$ with respect to a given state described by density operator $\hat{\rho}$

We express the state of the optical field $\hat{\rho}$ using Sudarshan's diagonal coherent state basis $i.e.$
 \begin{equation}
     \hat{\rho}=\int d^2 z {\vert z\rangle \langle
z\vert}\phi(z) \nonumber
 \end{equation}
where, $\phi(z)$ is a phase space distribution function or a probability distribution.\\
We find the following averages with respect to the density operator.
\begin{eqnarray}
(\Delta N)^2 &=& {\langle \hat{N}^2\rangle}-{\langle \hat{N}\rangle}^2 \nonumber\\
 &=& Tr\left(\hat{\rho}\hat{N}^2\right)-\left(Tr(\hat{\rho}\hat{N})\right)^2 
 \end{eqnarray}
 Now,
 \begin{eqnarray}
 Tr\left(\hat{\rho}\hat{N}^2\right) &=& Tr\left(\int d^2 z {|z\rangle \langle z|}\phi(z)\hat{N}^2\right)\nonumber\\
 &=& \int d^2 z\phi(z)\langle z|\hat{N}^2|z\rangle\nonumber\\
 &=& \int d^2 z\phi(z)\langle z|\hat{a}^{\dag}\hat{a}\hat{a}^{\dag}\hat{a}|z\rangle\nonumber\\
 &=& \int d^2 z\phi(z)|z|^2\langle z|(\hat{I}+\hat{a}^{\dag}\hat{a})|z\rangle \nonumber\\
 &=& \int d^2 z\phi(z)|z|^2 \left(1+|z|^2\right)
\end{eqnarray}
 Similarly,
 \begin{equation}
     \left(Tr(\hat{\rho}\hat{N})\right)^2=\left(\int d^2 z\phi(z)|z|^2\right)^2
 \end{equation}
 Therefore,
 \begin{eqnarray}
 (\Delta N)^2 &=& {\langle \hat{N}^2\rangle}-{\langle \hat{N}\rangle}^2 \nonumber\\
&=& \int d^2 z\phi(z)|z|^2 \left(1+|z|^2\right)-\left(\int d^2 z\phi(z)|z|^2\right)^{2}\nonumber\\
 \end{eqnarray}
\begin{align}
\therefore (\Delta N)^2-\langle \hat{N}\rangle = \int d^2 z\phi(z)|z|^2 
\left(1+|z|^2\right) - \nonumber\\ 
\left(\int d^2 z\phi(z)|z|^2\right)^{2} - \int d^2 z\phi(z)|z|^{2} \nonumber
\end{align}
\begin{align}
or, (\Delta N)^2-\langle \hat{N}\rangle = \int d^2 z\phi(z)|z|^2 \left(1 + |z|^2\right) - \nonumber\\ 
\left(\int d^2 z\phi(z)|z|^2\right)\left(\int d^2 z\phi(z)|z|^2\right) 
-\int d^2 z\phi(z)|z|^2\nonumber
\end{align}
\begin{eqnarray}
&=& \int d^2 z\phi(z)\left[|z|^4 +|z|^2 -|z|^2 \langle |z|^2 \rangle-|z|^2\right] \nonumber\\
&=&\int d^2 z\phi(z)\left[|z|^4 -2|z|^2 \langle |z|^2 \rangle+|z|^2\langle|z|^2\rangle\right]  \nonumber\\
&=&\int d^2 z\phi(z)\left[|z|^4 -2|z|^2 \langle |z|^2 \rangle\right]+\langle|z|^2\rangle\int d^2 z\phi(z) |z|^2\nonumber\\
&=&\int d^2 z\phi(z)\left[|z|^4 -2|z|^2 \langle |z|^2 \rangle\right]+{\langle|z|^2\rangle}^2\nonumber\\
&=&\int d^2 z\phi(z)\left[|z|^4 -2|z|^2 \langle |z|^2 \rangle\right]+{\langle|z|^2\rangle}^2\int d^2 z\phi(z)\nonumber\\ 
&=&\int d^2 z\phi(z)\left[|z|^4 -2|z|^2 \langle |z|^2 \rangle+{\langle|z|^2\rangle}^2\right]\nonumber\\
&=& \int d^2 z\phi(z)\left[|z|^2 - \langle|z|^2\rangle\right]^2 
\end{eqnarray}
In the above we used that $\phi(z)$ is normalized $i.e.$\\
$\int d^2 z \phi(z)=1$.
\begin{eqnarray}
    \therefore g^{(2)} (0) &=&  1+{\frac{(\Delta N)^2-{\langle \hat{N}\rangle}}{{\langle \hat{N}\rangle}^2}} \nonumber\\
    &=& 1+\frac{\int d^2 z\phi(z)\left[|z|^2 - \langle|z|^2\rangle\right]^2}{(\langle|z|^2\rangle)^2}
\end{eqnarray}
Now we compare the above expression with the second order coherence function that arises from classical description of electromagnetic field eq.(5.3) i.e. 

  $g^{(2)} (0) = 1+\frac{\int P(E)\left(|E|^2 -\langle |E|^2\rangle\right)^2 d^2 E}{(\langle |E|^2\rangle)^2}$

This appears similar in form to the quantum mechanical expression given in eq.(5.9) for $g^{(2)}(0)$ .
From the two expressions given above $P(E)$ is always positive for classical states; then $g^{(2)}(0)\geqslant 1$. This is allowed for classical states. There are no such classical fields for which $P(E)$ takes on negative values correspondingly $g^{(2)}(0)$ may be less than unity. On the other hand $\phi(z)>0$ occurs for classical states but $\phi(z)$ may take the form of more singular function than Dirac delta function or  is negative in some phase space points for non-classical states then $g^{(2)}(0)< 1$.\\
This unique photon anti-bunching property has no classical analogue. Hence, one can not infer it from classical coherence theory where probability function $P(E)$ takes always positive values. But this photon anti-bunching phenomena can be inferred using Sudarshan's quasi probability function $\phi(z)$.\\
For the most general quantum state $\hat{\rho}$, $\phi(z)$ is not a function in any ordinary mathematical sense, but a singular quantity, a so called tempered distribution of a particular class that can be precisely characterized. This result is truly fundamental to the theory of quantum optics, as this is the only way in which one can exhibit the clear distinction between classical and quantum nature of electromagnetic beams. States displaying photon anti-bunching or sub-poissonian photon statistics, squeezing and Hanbury-Brown-Twiss(HBT)  anti-correlations are truly quantum phenomena. These phenomena corresponds to singular, or at least non positive definite $\phi(z)$ according to Sudarshan's criterion and  can not be analyzed by classical optical theory. Now one can say that $\phi(z)$ is going beyond the usual classical probability function in considering all quantum states and shows that quantum and classical theories are different, the former overstepping the confined zone of the later. In fact Sudarshan's work was an attempt to express quantum optical coherence in the language of classical coherence-the quantum mechanical possibilities surpass classical boundaries.\\
The credit for formulating  and discovering the diagonal coherent state representation must go to E. C. G. Sudarshan. This work not to be dubbed as `Glauber's P representation' or `Glauber-Sudarshan representation'. In this context Prof. C. L. Mehta [11] said \emph{...It should be clear that there was and is only one diagonal representation, and Sudarshan had discovered it in its totality.} But unfortunately this result of Sudarshan did not receive the due credit and recognition. \\\\
\bibliographystyle{unsrt}

\vspace{0.5cm}
\textbf{Other References:}
\vspace{0.3cm}

[1] N. Mukunda, The Life and works of E. C. George Sudarshan, Resonance, February 2019; \\
https://doi.org/10.1007/s12045-019-0768-6.\\

[2] D. F. Walls, Evidence for the quantum nature of light, Nature, \textbf{280}, 451 (1979).\\

[3] Christopher Gerry and Peter Knight, Introductory Quantum Optics, Cambridge University Press (2004).

\end{document}